\documentclass{IEEEtaes}

\usepackage{color}
\usepackage{bm}
\usepackage{graphicx}
\usepackage{amsmath}
\usepackage{amssymb}
\usepackage{algorithm}
\usepackage{algorithmic}
\usepackage{multirow}
\usepackage{booktabs}
\usepackage{array}
\usepackage{lipsum}
\usepackage{enumerate}
\usepackage{stfloats}
\usepackage{subfigure}
\usepackage{cases}
\usepackage{diagbox}
\usepackage{cite}
\usepackage{algorithm,algorithmic}
\usepackage{balance}
\usepackage{makecell}

\usepackage{titlesec}

\titleformat{\subsubsection}[runin] 
{\normalfont\normalsize\itshape}
{\hspace{\parindent}\thesubsubsection)} 
{0.5em}                 
{}                     
[:~~]           

\titlespacing*{\subsubsection}
{0pt}    
{1ex plus .2ex minus .2ex} 
{0pt}

\newtheorem{remark}{Remark}
\newtheorem{prop}{Proposition}

\allowdisplaybreaks[4]

\begin{document}

	\title{Exploiting Target Location Distribution in MIMO Radar: PCRB vs. PSBP for Transmit Beampattern Design}

	\author{Lingyun Xu}
	\member{Graduate Student Member, IEEE}
	\affil{University of Electronic Science and Technology of China, Chengdu, China} 
	
	\author{Bowen Wang}
	\member{Graduate Student Member, IEEE}
	\affil{King’s College London, London, UK} 
	
	\author{Huiyong Li}
	\affil{University of Electronic Science and Technology of China, Chengdu, China}
	
	\author{Ziyang Cheng}
	\member{Senior Member, IEEE}
	\affil{University of Electronic Science and Technology of China, Chengdu, China}
	
	\receiveddate{This work was supported in part by the National Natural Science Foundation of China under Grants 62371096 and 62231006, and in part by the National Key Laboratory of Science and Technology on Space Microwave under Grant HTKJ2024KL504001.}
	
	\corresp{(\emph{Corresponding author: Ziyang Cheng})}
	
	\authoraddress{L. Xu, H. Li and Z. Cheng are with the School of Information and Communication Engineering, University of Electronic Science and Technology of China, Chengdu 611731, China. (email: xusherly@std.uestc.edu.cn, \{hyli, zycheng\}@uestc.edu.cn); B. Wang is with the Department of Engineering, King’s College London, London WC2R 2LS, U.K. (email: bowen.wang@kcl.ac.uk).}
	
	\maketitle

	\begin{abstract}This paper investigates the issue of how to exploit prior target location distribution for multiple input multiple output (MIMO) radar transmit beampattern design.
		We consider a MIMO radar aiming to estimate the random angular location parameters of a point target, whose prior distribution information can be exploited by the radar.
		First, we establish the models of the MIMO radar system and the target location distribution.
		Based on the considered models, we propose the first category of target location distribution exploitation methods by analyzing the radar direction-of-angle (DoA) estimation performance and deriving a general form of posterior Cram{\'e}r-Rao bound (PCRB) as the lower bound of the mean square error of DoA estimation.
		Following this, to investigate the potential of leveraging prior information from two distinct perspectives, thereby providing deeper insights into the estimation process, we proposed the second category of target location distribution exploitation methods by introducing a novel radar metric, probability scaled beampattern (PSBP), from the perspective of radar beampattern.
		To compare the two methods, we formulate the PCRB and PSBP oriented radar transmit beampattern design problems and propose corresponding low-complexity and convergence-guaranteed algorithms to tackle them.
		Finally, numerical simulations are conducted in different scenarios to provide a comprehensive evaluation and comparison of the radar performance.
	\end{abstract}

	\vspace{-1em}
	\section{Introduction}
	Multiple input multiple output (MIMO) radars employ multiple antennas at both transmitter and receiver, 
	which can transmit multiple probing signals and receive multiple target echoes \cite{fishler2004mimo, donnet2006mimo, li2008mimo, li2007mimo}.
	Compared with the traditional standard phase array radars, MIMO radars offer better accuracy of angular estimation and suppression of clutter/interference with a higher degree of freedom (DoF) \cite{fishler2004performance,li2007mimo}.
	With these advantages, MIMO radars are promising in many applications, i.e., electronic reconnaissance \cite{tang2011new}, geophysical monitoring \cite{mugnai2022multiple}, medical imaging \cite{bliss2006mimo}, etc.
	Driven by this, the investigations of MIMO radars have been widely carried out by many researchers.
	
	Among the investigations of MIMO radars, the higher DoF provided by MIMO radar allows for flexible transmit beampattern design with proper properties, enhancing MIMO radar performance. 
	Therefore, some works have focused on the MIMO radar transmit beampattern design with a desirable transmit beampattern \cite{cheng2017constant, fan2018constant,fan2019mimo,fan2021spectrally,gong2024frequency,gong2014transmit}.
	For example, authors in \cite{cheng2017constant} proposed a MIMO radar waveform design with the constant modulus constraint, where two methods were devised to minimize the square error and absolute error between the designed beampattern and the desired one, respectively.
	Another work in \cite{fan2019mimo} proposed a MIMO radar waveform design to achieve quasi-equiripple beampattern, where a new weighted $l_p$-norm matching metric was introduced and the matching error between the designed beampattern and the desired one was minimized.
	Moreover, the authors in \cite{fan2019mimo} also extended the MIMO radar waveform design to the wideband in a spectrally dense environment in \cite{fan2021spectrally}, where two design problems were formulated with different practical considerations by minimizing the beampattern matching error and peak sidelobe level (PSL), respectively.
	Furthermore, authors in \cite{gong2024frequency} considered the waveform design for wideband MIMO radar, where frequency-invariant beampattern was optimized subject to nulling and peak-to-average-power ratio (PAPR) constraint.
	To realize a fairer control of both mainlobe and sidelobe levels, two methods for a constant modulus MIMO radar waveform design were proposed in \cite{fan2018constant}, where the first one was to maximize the ratio of the minimum mainlobe level to the PSL, and the second one was to minimize the PSL subject to the mainlobe ripple constraint.

	Compared with waveform design with the desirable transmit beampattern, some more quantifiable metrics, i.e., signal-to-interference-plus-noise ratio (SINR) \cite{cui2013mimo,tang2016joint,cheng2018spectrally,yu2020mimo,gong2022joint}, mutual information, relative entropy \cite{tang2010mimo,naghsh2017information,sen2010ofdm}, etc., are also widely considered in MIMO radar design, which affects radar detection performance.
	For instance, authors in \cite{cui2013mimo} considered a MIMO radar waveform design problem in the presence of interference, where the output SINR was maximized while meeting the constant modulus and similarity constraints.
	In \cite{cheng2018spectrally}, a spectrally compatible MIMO radar waveform design was proposed in the presence of multiple targets, where the waveform energy was minimized while satisfying the requirement of individual SINR for each target.
	Based on information theory, authors in \cite{tang2010mimo} proposed the optimal MIMO radar waveform in colored noise, where two criteria were used: the mutual information between target impulse response and target echoes and the relative entropy between two hypotheses were maximized, respectively.
	
	When focusing more intently on the problem of radar target localization in MIMO radar systems, a standard mathematical metric for measuring the accuracy of radar estimation of target parameters, Cram{\'e}r-Rao bound (CRB) \cite{bekkerman2006target, tang2013maximum,boyer2009co}, is considered in MIMO radar design.
	In particular, the application of CRB is crucial in the estimation of direction-of-arrival (DoA) in MIMO radar systems.
	By calculating CRB, we can evaluate to what extent can the DoA estimation performance of different waveform design methods reach the theoretical lower bound \cite{hassanien2011transmit}.
	For example, a CRB-based study for MIMO radar waveform optimization was carried out in \cite{li2007range}, where the waveform design problem was formulated by minimizing the CRB of estimation for multiple targets in the presence of spatially colored interference and noises. 
	Moreover, authors in \cite{wang2011joint} proposed a joint design of MIMO radar waveform and biased estimator in the presence of signal-dependent noises, where a newly introduced constrained biased CRB was minimized to enhance the estimation performance.
	Another waveform design for MIMO radars to realize high-resolution localization was proposed in \cite{wu2023waveform}, where the asymptotic mean square error (MSE) of DoA estimation with multiple signal classification (MUSIC) algorithms was minimized to close to the CRB.

	The aforementioned works \cite{bekkerman2006target, tang2013maximum,boyer2009co,hassanien2011transmit,li2007range,wang2011joint,wu2023waveform} considered the time-invariant systems, where the location parameters of targets were assumed to be deterministic.
	However, in the time-varying systems, the location parameters of targets should be considered random, where the standard CRB is not applicable.
	In this case, a lower bound of random parameter estimation analogous to the CRB derived in \cite{van1968detection}, called posterior CRB (PCRB), also known as Bayesian CRB (BCRB), becomes a more appropriate metric \cite{tichavsky1995posterior,tichavsky1998posterior}.
	By applying the PCRB, which integrates both distribution information of the random parameters and observed data, we can achieve a more accurate and comprehensive assessment of the estimation accuracy for random target parameters.
	To this end, there have been many researches using PCRB to design radar waveforms when the target location parameters are random \cite{hurtado2008adaptive,huleihel2013optimal,sharaga2014optimal,sharaga2015optimal,godrich2010target,chavali2010cognitive}.
	Although these works \cite{hurtado2008adaptive,huleihel2013optimal,sharaga2014optimal,sharaga2015optimal,godrich2010target,chavali2010cognitive} realized satisfactory DoA estimation performance in certain scenarios, they still have three main limitations: 
	\textit{i)} They did not consider the prior target location distribution information, simply assumed a uniform distribution among the whole range, and recursively inferred a posterior PDF based on the previous observations and estimations. 
	However, the estimation errors are unavoidable and would be accumulated during the recursion, leading to an inaccurate approximated posterior PDF and thus degradation in DoA estimation performance.
	\textit{ii)} They required uninterrupted measurements to compute recursive equations to obtain the PCRB of estimation MSE, leading to an increasing computational complexity over time.
	\textit{iii)} They only designed radar transmit beampattern by minimizing the PCRB, but PCRB is a theoretical lower bound of estimation MSE.
	Whether such waveforms can achieve the MSE close to the PCRB, and whether there is any other method that can achieve better estimation performance, are still open questions.
	
	Motivated by the facts mentioned above, we investigate the problem of MIMO radar transmit beampattern designed by exploiting target location distribution.
	Specifically, the main contributions of this work can be summarized as follows.
	\vspace{-1em}
	\begin{itemize}
		\item \textit{Different target location distribution models:} 
		Considering practical radar sensing scenarios, we establish two typical target angular location distribution models based on the prior statistical characteristics of the target location.
		Specifically, the first scenario is that the target is located on certain angular intervals with high occurrence probabilities, where the target location distribution is characterized by the uniform mixture model.
		The second scenario is that the target is located around certain angular points with high occurrence probabilities, where the target location distribution is characterized by the truncated Gaussian mixture model.
		
		\item \textit{PCRB oriented transmit beampattern} optimization method:
		Based on the considered system and target location distribution models, we analyze the MSE of DoA estimation and derive a general form of PCRB as the lower bound of the MSE, which is dependent on the target location distribution probability density function (PDF) and radar waveform.
		Guided by this, we formulate a PCRB oriented transmit beampattern design problem, which can be solved by the proposed alternating direction method of multipliers (ADMM) based algorithm.
		
		\item \textit{PSBP oriented transmit beampattern} optimization method:
		To further explore the potential of exploiting target location distribution, thereby providing more insights into the estimation, we propose exploiting the target location distribution from the radar beampattern perspective.
		Specifically, we introduce a novel metric, probability scaled beampattern (PSBP), which is also dependent on the target location distribution PDF.
		For different requirements of desirable beampattern, we formulate two PSBP oriented transmit beampattern design problems: fair PSBP design and integrated PSBP design, which can be also solved by the proposed ADMM-based algorithms.
		
		\item \textit{Multi-scenario radar sensing performance analysis:} 
		We provide extensive numerical simulations in different scenarios to provide a comprehensive performance evaluation and comparison of the proposed transmit beampattern design methods.
		
	\end{itemize}
	
	\textit{Organization:}
	Section II introduces the system model.
	Section III presents the target location distribution exploitation, gives the performance analysis, and formulates problems.
	Section IV and Section V derive the solutions to PCRB and PSBP oriented transmit beampattern design problems, respectively.
	Section VI demonstrates the numerical simulations and Section VII concludes this work.
	
	\textit{Notations:} This paper uses lower-case letters $a$ for scalars, lower-case bold letters $\mathbf{a}$ for vectors and upper-case bold letters $\mathbf{A}$ for matrices. For a matrix $\mathbf{A}$, the element in the $i$-th row and the $j$-th column is denoted by ${\bf{A}}(i,j)$.  $\mathbb{C}^{n}$ and $\mathbb{C}^{m\times n}$ denote an $n$ dimensional complex-valued vector and an $m$ by $n$ dimensional complex-valued space, respectively. $(\cdot)^T$ and $(\cdot)^H$ denote transpose and conjugate transpose operators, respectively. $\left|  \cdot  \right|$ represents a determinant or absolute value up to the context. ${\left\|  \cdot  \right\|_F}$ and ${\rm Tr}({\cdot})$ denote the Frobenius norm and trace, respectively. ${\mathbb E}\left\{   \cdot \right\}$ denotes expectation. $\Re \left\{  \cdot  \right\}$  and $\Im \left\{  \cdot  \right\}$ denote the real part and imaginary part of a complex-valued number, respectively. $\mathcal{C}\mathcal{N}\left( {0,{\mathbf{R}}} \right)$ denotes the zero-mean complex Gaussian distribution with covariance matrix ${\mathbf{R}}$.
	
	\vspace{-1em}
	\section{System Model}
	
	\subsection{Signal Model}
	
	\begin{figure}[!t]
		\centering  \includegraphics[width=1\linewidth]{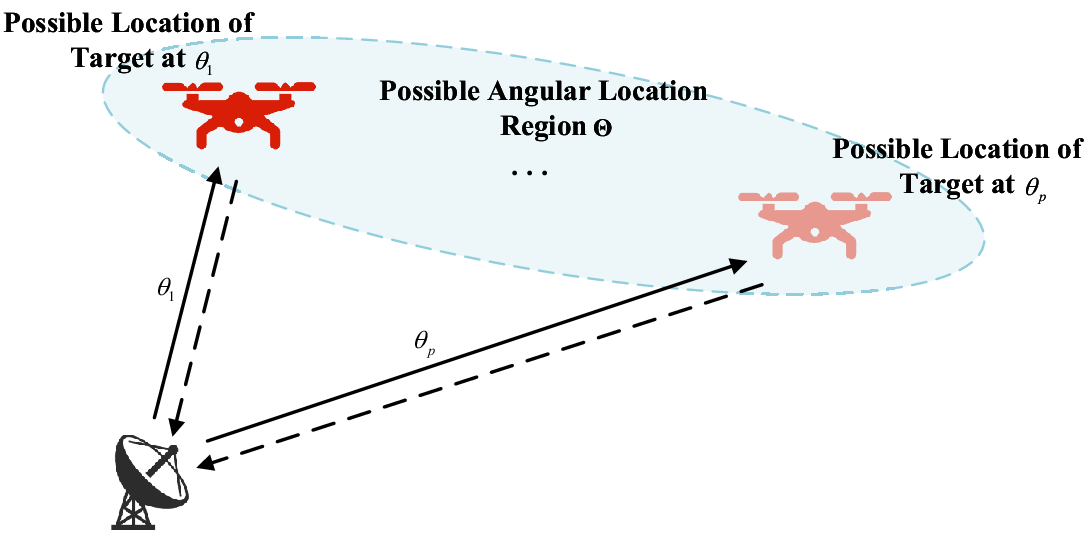}
		\vspace{-1.5em}
		\caption{Illustration of target localization with angular location distribution information available at the MIMO radar.}
		\label{fig:scene_graph}
		\vspace{-1.5em}
	\end{figure}

	As shown in Fig. \ref{fig:scene_graph}, we consider a colocated MIMO radar, where the radar transmitter probes waveforms to illuminate targets, and the radar receiver collects the reflected echoes to conduct estimation.
	{Unlike most existing papers \cite{bekkerman2006target, tang2013maximum,boyer2009co,hassanien2011transmit,li2007range,wang2011joint,wu2023waveform,hurtado2008adaptive,huleihel2013optimal,sharaga2014optimal,sharaga2015optimal,godrich2010target,chavali2010cognitive}, we assume that the exact angular location of the target is \textit{random} and its prior distribution information is available at the radar and can be exploited to enhance sensing performance.}
	The colocated radar is equipped with $M_t$ transmit antennas and $M_r$ receive antennas, where the antenna arrays in both transmitter and receiver are assumed to be uniform linear arrays (ULAs) with half-wavelength element spacing.
	The discrete-time waveform radiated by each antenna is represented by ${{{\bf{x}}_m}\left( l \right)}, m = 1, \dots, M_t, l = 1, \dots, L$, where $L$ denotes the number of samples for each radar pulse.
	Let ${\bf{x}}\left( l \right) = {\left[ {{{\bf{x}}_1}\left( l \right), \ldots ,{{\bf{x}}_{{M_t}}}\left( l \right)} \right]^T} \in {\mathbb C^{{M_t} }}$ be the transmit waveform vector at the $l$-th sample of the $M_t$ transmit antennas.
	Thus, the received signal at the $l$-th sample, ${\bf{y}}\left( l \right) \in {{\mathbb C}^{{M_r}}}$, can be expressed as
	\begin{equation}\label{eq:1}
	{\bf{y}}\left( l \right) = \varsigma {{\bf{a}}_r}\left( \theta  \right){\bf{a}}_t^H\left( \theta  \right){\bf{x}}\left( l \right) + {\bf{z}}\left( l \right) ,
	\end{equation}
	where $\varsigma $ denotes the complex amplitude of the target, which is assumed to be a deterministic parameter, ${\bf{z}}\left( l \right) \sim {\cal C}{\cal N}\left( {0,\sigma _{\rm{r}}^2{{\bf{I}}_{{M_r}}}} \right)$ denotes the additive white Gaussian noise (AWGN).
	The transmit and receive steering vectors are ${{\bf{a}}_t}(\theta ) = {[ {{e^{ - \jmath \pi {{\frac{{M_t} - 1}{2}}}\sin \theta }}, \ldots ,{e^{\jmath \pi {{\frac{{M_t} - 1}{2}}}\sin \theta }}} ]^T} \in {\mathbb{C}^{{M_t}}}$ and ${{\bf{a}}_r}(\theta ) = {[ {{e^{ - \jmath \pi {{\frac{{M_r} - 1}{2}}}\sin \theta }}, \ldots ,{e^{\jmath \pi {{\frac{{M_r} - 1}{2}}}\sin \theta }}} ]^T} \in {\mathbb{C}^{{M_r}}}$, respectively.

	By collecting $L$ samples of the received signal, the received signal during one frame, ${\bf Y} \in {{\mathbb C}^{{M_r} \times L}}$, can be given by
	\begin{equation}\label{eq:2}
	{\bf{Y}} = \left[ {{\bf{y}}\left( 1 \right), \ldots ,{\bf{y}}\left( L \right)} \right] = \varsigma {\bf{A}}\left( \theta  \right){\bf{X}} + {\bf{Z}} ,
	\end{equation}
	where ${\bf{X}} = \left[ {{\bf{x}}\left( 1 \right), \ldots ,{\bf{x}}\left( L \right)} \right] \in {\mathbb C^{{M_t} \times L}}$ is the space-time transmit waveform matrix, and we define ${\bf{A}}\left( \theta  \right) = {{\bf{a}}_r}\left( \theta  \right){\bf{a}}_t^H\left( \theta  \right)$ and ${\bf{Z}} = \left[ {{\bf{z}}\left( 1 \right), \ldots ,{\bf{z}}\left( L \right)} \right]$.
	
	The radar transmitter typically exhibits a large dynamic range. 
	To avoid out-of-band radiation and signal distortion, the peak-to-average-power ratio (PAPR) of the radar transmit waveforms should be properly limited \cite{stoica2008waveform,de2011design,cheng2017mimo}. 
	To this end, we define the PAPR as follows
	\begin{equation}\label{eq:3}
	{\rm{PAPR}}\left( {\bf{X}} \right) = \frac{{\mathop {\max }\limits_{m,l} {{\left| {{\bf{X}}\left( {m,l} \right)} \right|}^2}}}{{\frac{1}{{{M_t}L}}\left\| {\bf{X}} \right\|_F^2}} \le \kappa , \forall m,l ,
	\end{equation}
	where $\kappa \ge 1$ denotes the PAPR threshold of the radar system.
	Without loss of generality, we assume that the total transmit power is $P$, i.e., $\left\| {\bf{X}} \right\|_F^2 = {P}$.
	By substituting the transmit power budget to \eqref{eq:3}, we can rewrite the PAPR constraint into
	\begin{equation}
	\begin{aligned}
	\mathrm{PAPR}(\mathbf{X}) \to \left\{ 
	\begin{array}{l}
	\left\| {\bf{X}} \right\|_F^2 = {P} , \\
	\frac{{{M_t}L}}{{ P}}\mathop {\max }\limits_{m,l} {\left| {{\bf{X}}\left( {m,l} \right)} \right|^2}  \le \kappa,  \forall m,l .
	\end{array}
	\right.
	\end{aligned}
	\end{equation}
	
	\vspace{-1em}
	\subsection{Target Location Distribution Model}\label{Sec:2-B}
	
	As shown in Fig. \ref{fig:scene_graph}, a MIMO radar aims at estimating the DoA of a point target.
	It is assumed that the exact angle $\theta  \in {\bf{\Theta }} = \left[ { - \pi /2,\pi /2} \right]$ of the target is a \textit{random} parameter, whose prior location distribution information can be extracted from historical observations and empirical data is available at the radar.
	Let $f\left( \theta_p  \right)$ denote the target angular location PDF at $\theta_p$.
	In the following, based on the statistical characteristics of the angular location $\theta$ and the considered practical scenarios, we proceed with modeling the continuous PDF of $\theta$ as $f\left( \theta  \right)$, which leads to the following two categories.

	\subsubsection{Distribution Model 1}
	\textit{High Occurrence Probability on Certain Angular Intervals.}
	Suppose the prior target angular location distribution available at the radar is that the potential target may be located on $K$ angular intervals, $ {{{\bf{\Theta }}_1}, \ldots ,{{\bf{\Theta }}_K}}$.
	
	We assume that the probability of the target appearing on the $k$-th possible angular location interval is ${p_k}$ and $\sum\nolimits_{k = 1}^K {p_k}  = 1$, where the target angle obeys a uniform distribution on each possible angular location interval.
	The continuous PDF of the $k$-th possible angular location interval is then given by
	\begin{equation}\label{eq:16}
	{f_k}\left( \theta  \right) = {p_k}\frac{1}{{{\theta _{k2}} - {\theta _{k1}}}},\theta  \in {{\bf{\Theta }}_k},
	\end{equation}
	where ${\theta _{k1}}$ and ${\theta _{k2}}$ denote the left and right end points of the $k$-th possible angular location interval ${{\bf{\Theta }}_k}$.
	
	Based on this model, we can characterize the target angular location distribution by defining the continuous mixed uniform PDF as
	\begin{equation}\label{eq:17}
	f\left( \theta  \right) = \textstyle\sum\limits_{k = 1}^K {{f_k}\left( \theta  \right)}\left[ {u\left( {\theta  - {\theta _{k1}}} \right) - u\left( {\theta  - {\theta _{k2}}} \right)} \right],
	\end{equation}
	which is the sum of the probability-weighted $K$ PDFs of the uniform distribution on possible target angular location distribution intervals.
	Besides, $u(\cdot)$ represent the unit step function.
	
	\subsubsection{Distribution Model 2} 
	\textit{High Occurrence Probability at Certain Angular Points}.
	Then, we consider another typical scenario, where the prior target location distribution information available at the radar is that the potential target may be located at $K$ angular points, 
	$\theta_1,\dots,\theta_K$, with a small uncertainty.
	We assume that the probability of the target appearing around the $k$-th point is ${{p_k}}$ and $\sum\nolimits_{k = 1}^K {p_k}  = 1$.
	Since the exact angle lies within the bounded angular range $[- \pi/2,\pi/2]$, the target angle obeys a truncated Gaussian distribution \cite{burkardt2014truncated} around each possible angular point.
	Accordingly, the continuous PDF around the $k$-th possible angular point is modeled as
	\begin{equation}\label{eq:19}
	{f_k}\left( \theta  \right) = p_k \frac{{\phi \left( {{\theta _k},\sigma _\theta ^2;\theta } \right)}}{{\Phi \left( {{\theta _k},\sigma _\theta ^2;\pi /2} \right) - {\Phi\left( {{\theta _k},\sigma _\theta ^2; - \pi /2} \right)}}},\theta  \in {\bf{\Theta }},
	\end{equation}
	where ${\phi ( {{\theta _k},\sigma _\theta ^2;\theta } )}$ denotes the Gaussian PDF with mean $\theta_k$ and variance $\sigma_{\theta}^2$, and $\Phi ( {{\theta _k},\sigma _\theta ^2;\tilde \theta } )$ denotes its corresponding cumulative distribution function (CDF) at $\tilde \theta$.
	
	Based on this model, we can characterize the target angular location distribution by defining the continuous mixed truncated Gaussian PDF as
	\begin{equation}\label{eq:20}
	f\left( \theta  \right) = \textstyle\sum\limits_{k = 1}^K {{f_k}\left( \theta  \right)} \left[ {u\left( {\theta  + \pi /2} \right) - u\left( {\theta  - \pi /2} \right)} \right],
	\end{equation}
	which is the sum of the probability-weighted $K$ PDFs of the truncated Gaussian distribution around possible target angular location points.
	
	In this paper, we aim to discuss how to effectively use target prior location distribution information to design radar waveforms that improve radar estimation performance. 
	Thus, in the following sections, we will discuss the exploitation of target location distribution from the perspectives of the PCRB and beampattern design.
	
	\vspace{-1em}
	\section{Target Location Distribution Exploitation and Performance Analysis}
	
	In this section, we analyze radar DoA estimation performance using the MSE estimator, yielding a novel PCRB-based prior target location distribution exploitation method. 
	Then, we propose exploiting prior target location distribution from the beampattern perspective and derive a novel PSBP metric.

	\subsection{PCRB-based Target Location Distribution Exploitation}
	
	Since we focus on location estimation, it is straightforward to evaluate the estimation performance from the PCRB perspective \cite{tichavsky1995posterior,tichavsky1998posterior}. 
	Given that we assume the target location distribution is available, we will focus on the PCRB analysis \cite{tichavsky1995posterior,tichavsky1998posterior}. 
	Because we consider a general $ f(\theta) $ with two typical distribution models, in the following sections, we will formulate the general form of PCRB of the considered radar system and apply the general results to these two typical cases.
	
	\subsubsection{General Form of PCRB}
	
	For notations, we can define the collection of the target's random parameter as ${\bm{\omega }} = {[ {\theta ,{\varsigma _{\rm{R}}},{\varsigma _{\rm{I}}}} ]^T}$, where ${\varsigma _{\rm{R}}} = \Re \{ \varsigma  \}$ and ${\varsigma _{\rm{I}}} = \Im \{ \varsigma \}$.
	It is assumed that $\theta$ and $\varsigma$ are independent of each other.
	
	We suppose that ${{\bf{\hat {\bm \omega} }}}$ is the unbiased estimation of ${{\bm{\omega }}}$ and then the estimation MSE of  ${{\bm{\omega }}}$ is ${\rm{MSE}_{\bm{\omega }}}\left( {{\bm{\hat \omega }}} \right) = {\mathbb E}\{ {\| {{\bm{\omega }} - {\bm{\hat \omega }}}\|_F^2} \}$.
	Since the MSE is equal to the generalized error variance for an unbiased estimator \cite{anderson1958introduction} and the information of ${\bm{\omega }}$ can be exploited, we define PCRB as the lower bound of the MSE
	\begin{equation}\label{eq:5}
	{\rm{MS}}{{\rm{E}}_{\bm{\omega }}}\left( {{\bm{\hat \omega }}} \right) \ge {\rm{PCRB}}_ {{\bm{\omega }}|{\bf{Y}}} = \left| {{\bf{F}}_{{\bm{\omega }}|{\bf{Y}}}^{ - 1}} \right| , 
	\end{equation}
	where ${{\bf{F}}_{{\bm{\omega }}|{\bf{Y}}}} \in {{\mathbb C}^{3 \times 3}}$  denotes the posterior Fisher information matrix (FIM) for the estimation of ${\bm{\omega }}$, which is given by
	\begin{equation}\label{eq:6}
	{{\bf{F}}_{{\bm{\omega }}|{\bf{Y}}}} = {{\bf{F}}_{\rm{S}}} + {{\bf{F}}_{\rm{P}}} ,
	\end{equation}
	where ${{\bf{F}}_{\rm{S}}}$ denotes the FIM extracted from received signals, and ${{\bf{F}}_{\rm{P}}}$ denotes the FIM extracted from distribution information.
	Specifically, $\mathbf{F}_\mathrm{S}$ and $\mathbf{F}_\mathrm{P}$ are given by
	\begin{subequations}\label{eq:7}
		\begin{align}
		&{{\bf{F}}_{\rm{S}}} = {\mathbb E} \left\{ {\frac{{\partial \ln \left( {f\left( {{\bf{Y}}|{\bm{\omega }}} \right)} \right)}}{{\partial {\bm{\omega }}}}{{\left( {\frac{{\partial \ln \left( {f\left( {{\bf{Y}}|{\bm{\omega }}} \right)} \right)}}{{\partial {\bm{\omega }}}}} \right)^H}}} \right\} , \label{eq:7a}\\
		&{{\bf{F}}_{\rm{P}}} =  {\mathbb E}\left\{ {\frac{{\partial \ln f\left( {\bm{\omega }} \right)}}{{\partial {\bm{\omega }}}}{{\left( {\frac{{\partial \ln f\left( {\bm{\omega }} \right)}}{{\partial {\bm{\omega }}}}} \right)^H}}} \right\} , \label{eq:7b}
		\end{align}
	\end{subequations}
	where ${f\left( {{\bf{Y}}|{\bm{\omega }}} \right)}$ denotes the conditional PDF of $\bf Y$ with a given ${\bm{\omega }}$ and $f\left( {\bm{\omega }} \right)$ denotes the PDF of ${\bm{\omega }}$.
	Accordingly, we can divide the derivation of the posterior FIM \eqref{eq:6} into two steps, the derivations of ${\bf F}_{\rm S}$ and ${\bf F}_{\rm P}$.
	
	\textbf{Step 1: The Derivation of ${\bf F}_{\rm S}$.}
	We deal with the FIM extracted from the received signal, ${{\bf{F}}_{\rm{S}}}$, as follows.
	The joint conditional distribution of ${\bf{Y}} = \left[ {{\bf{y}}\left( 1 \right), \ldots ,{\bf{y}}\left( L \right)} \right]$ is given by
	\begin{equation}\label{eq:8}
	f\left( {{\bf{Y}}|{\bm{\omega }}} \right) = \frac{1}{{{\pi ^{{M_r}L}}\left| {\sigma _{\rm{r}}^2{{\bf{I}}_{{M_r}L}}} \right|}}\exp \left\{ { - \frac{{\left\| {{\bf{Y}} - \varsigma {\bf{A} \left( \theta  \right){\bf X}}} \right\|_F^2}}{{\sigma _{\rm{r}}^2}}} \right\}.
	\end{equation}
	Accordingly, the log-likelihood function of $\bf Y$ given ${\bm{\omega }}$ can be calculated as
	\begin{equation}\label{eq:9}
	\begin{aligned}
	l\left( {{\bf{Y}}|{\bm{\omega }}} \right) =& \ln f\left( {{\bf{Y}}|{\bm{\omega }}} \right) \\
	= &- {M_r}L\ln \left( {\pi \sigma _{\rm{r}}^2} \right) - \frac{{\left\| {\bf{Y}} \right\|_F^2}}{{\sigma _{\rm{r}}^2}} - \frac{{{{\left| \varsigma  \right|}^2}\left\| {{\bf{A}}\left( \theta  \right){\bf{X}}} \right\|_F^2}}{{\sigma _{\rm{r}}^2}}\\
	&+ \frac{{2\Re \left\{ {{\rm{Tr}}\left\{ {{\varsigma ^*}{{\bf{X}}^H}{{\bf{A}}^H}\left( \theta  \right){\bf{Y}}} \right\}} \right\}}}{{\sigma _{\rm{r}}^2}}.
	\end{aligned}
	\end{equation}
	Then, we have the following proposition to calculate ${{\bf{F}}_{\rm{S}}}$.
	\begin{prop}\label{prop:1}
		Based on \eqref{eq:7a} and \eqref{eq:9}, ${{\bf{F}}_{\rm{S}}}$ can be decomposed into blocks, which is given by
		\begin{equation}\label{eq:10}
		{{\bf{F}}_{\rm{S}}} = \left[ {\begin{array}{*{20}{c}}
			{{F_{\theta \theta }}}&{{{\bf{F}}_{\theta \varsigma }}}\\
			{{{\bf{F}}_{\varsigma \theta }}}&{{{\bf{F}}_{\varsigma \varsigma }}}
			\end{array}} \right] \in {\mathbb{C}^{3 \times 3}} ,
		\end{equation}
		where
		\begin{subequations}
			\begin{align}
			{F_{\theta \theta }} =& \frac{{2{{\left| \varsigma  \right|}^2}}}{{\sigma _{\rm{r}}^2}}{\rm{Tr}}\left\{ {{{\bf{X}}^H}{{\bf{\Xi }}_1}{\bf{X}}} \right\} , \\
			{{\bf{F}}_{\theta \varsigma }} =& \frac{2}{{\sigma _{\rm{r}}^2}}{\rm{Tr}}\left\{ {{{\bf{X}}^H}{{\bf{\Xi }}_2}{\bf{X}}} \right\}\left[ {{\varsigma _{\rm{R}}},{\varsigma _{\rm{I}}}} \right] , \\
			{{\bf{F}}_{\varsigma \theta }} =& {\bf{F}}_{\theta \varsigma }^H = {\left( {\frac{2}{{\sigma _{\rm{r}}^2}}{\rm{Tr}}\left\{ {{{\bf{X}}^H}{{\bf{\Xi }}_2}{\bf{X}}} \right\}\left[ {{\varsigma _{\rm{R}}},{\varsigma _{\rm{I}}}} \right]} \right)^H} , \\
			{{\bf{F}}_{\varsigma \varsigma }} =& \frac{2}{{\sigma _{\rm{r}}^2}}{\rm{Tr}}\left\{ {{{\bf{X}}^H}{{\bf{\Xi }}_3}{\bf{X}}} \right\}{{\bf{I}}_2}  , \label{eq:11}
			\end{align}
		\end{subequations}
		with\footnote{ {To be consistent with the existing works \cite{hurtado2008adaptive,huleihel2013optimal,sharaga2014optimal,sharaga2015optimal,godrich2010target,chavali2010cognitive}, the expectations involved in this paper are computed numerically using Monte-Carlo (MC) integration.}}
		\begin{subequations}
			\begin{align}
			{{\bf{\Xi }}_1} & = \textstyle\int_{ - \pi /2}^{\pi /2} {f\left( \theta  \right)} \left\| {{{{\bf{\dot a}}}_r}\left( \theta  \right)} \right\|_F^2{{\bf{a}}_t}\left( \theta  \right){\bf{a}}_t^H\left( \theta  \right) d\theta + {{\bf{\Xi }}_2} , \label{neq:3-16a}\\
			{{\bf{\Xi }}_2} & = {M_r}\textstyle\int_{ - \pi /2}^{\pi /2} {f\left( \theta  \right){{{\bf{\dot a}}}_t}\left( \theta  \right){\bf{a}}_t^H\left( \theta  \right)d\theta } , \label{neq:3-16b} \\
			{{\bf{\Xi }}_3} & = {M_r}\textstyle\int_{ - \pi /2}^{\pi /2} {f\left( \theta  \right){{\bf{a}}_t}\left( \theta  \right){\bf{a}}_t^H\left( \theta  \right)d\theta }, \label{neq:3-16c}
			\end{align}
		\end{subequations}
		${\bf{\dot a}\left( \theta\right)}$ denoting the derivative of ${\bf{ a}\left( \theta\right)}$ about $\theta$.
	\end{prop}
	\begin{IEEEproof}
		Please refer to Appendix \ref{app:A}.
	\end{IEEEproof}
	
	\textbf{Step 2: The Derivation of ${\bf F}_{\rm P}$.}
	We calculate the FIM extracted from the distribution, ${{\bf{F}}_{\rm{P}}}$.
	Since the angle $\theta$ is a random parameter, and complex amplitude $\varsigma$ is a deterministic parameter, we can derive 
	\begin{equation}\label{eq:12}
	\frac{{\partial \ln f\left( {\bm{\omega }} \right)}}{{\partial {\bm{\omega }}}} = {\left[ {\frac{{\partial \ln f\left( \theta  \right)}}{{\partial \theta }},0,0} \right]^T} .
	\end{equation}
	
	Based on \eqref{eq:7b} and \eqref{eq:12}, ${{\bf{F}}_{\rm{P}}}$ can be calculated and decomposed into
	\begin{equation}\label{eq:13}
	{{\bf{F}}_{\rm{P}}} = \left[ {\begin{array}{*{20}{c}}
		{B_{\theta \theta }} &0&0\\
		0&0&0\\
		0&0&0
		\end{array}} \right]\in {\mathbb{C}}^{3\times3} , 
	\end{equation}%
	where 
	\begin{equation}\label{neq:19}
	{B_{\theta \theta }} = \textstyle\int_{ - \pi /2}^{\pi /2} {f\left( \theta  \right){{\left( {\frac{{\partial \ln f\left( \theta  \right)}}{{\partial \theta }}} \right)}^2}d\theta }  \buildrel \Delta \over = \Lambda
	\end{equation}

	Based on the above two-step formulations, the PCRB in \eqref{eq:5} can be expressed as
	\begin{equation}\label{eq:14}
	{\rm{PCR}}{{\rm{B}}_{{\bm{\omega }}|{\bf{Y}}}} = \left| {{{\left[ {\begin{array}{*{20}{c}}
					{{F_{\theta \theta }} + {B_{\theta \theta }}}&{{{\bf{F}}_{\theta \varsigma }}}\\
					{{{\bf{F}}_{\varsigma \theta }}}&{{{\bf{F}}_{\varsigma \varsigma }}}
					\end{array}} \right]}^{ - 1}}} \right| .
	\end{equation}
	To characterize the DoA estimation performance of the MIMO radar, we derive the PCRB for the estimation of $\theta$ with unknown $\varsigma$, ${\rm{PCR}}{{\rm{B}}_\theta }\left( {\bf{X}} \right)$, by extracting the upper-left block corresponding to $\theta$ \cite{van2002optimum}, which is given by
	\begin{equation}\label{eq:15}
	\begin{aligned}
	&{\rm{PCR}}{{\rm{B}}_\theta }\left( {\bf{X}} \right) \\
	&={\left[ {{F_{\theta \theta }} + {B_{\theta \theta }} - {{\bf{F}}_{\theta \varsigma }}{\bf{F}}_{\varsigma \varsigma }^{ - 1}{{\bf{F}}_{\theta \varsigma }^H}} \right]^{ - 1}}\\
	&= {\left[ {\frac{{2{{\left| \varsigma  \right|}^2}}}{{\sigma _{\rm{r}}^2}}\left({\rm{Tr}}\left\{ {{{\bf{X}}^H}{{\bf{\Xi }}_1}{\bf{X}}} \right\}   - \frac{{{{\left| {{\rm{Tr}}\left\{ {{{\bf{X}}^H}{{\bf{\Xi }}_2}{\bf{X}}} \right\}} \right|}^2}}}{{{\rm{Tr}}\left\{ {{{\bf{X}}^H}{{\bf{\Xi }}_3}{\bf{X}}} \right\}}} \right)+ \Lambda } \right]^{ - 1}}
	\end{aligned}
	\end{equation}
	
	\subsubsection{PCRB for Typical Distribution Model}
	
	Based on \eqref{eq:15}, the PCRB is a function of the radar waveform $\bf X$ with some distribution-related parameters, i.e., $\mathbf{\Xi}_1$, $\mathbf{\Xi}_2$, $\mathbf{\Xi}_3$, and $\Lambda$.
	For the specific distribution model, the distribution-related parameters can be calculated by substituting the prior PDF of angular location to the equations \eqref{neq:3-16a}-\eqref{neq:3-16c} and \eqref{neq:19}.
	Therefore, the PCRB for the two typical distribution models considered in Sec. \ref{Sec:2-B} can be calculated by substituting the continuous mixed uniform PDF \eqref{eq:17} and the continuous mixed truncated Gaussian PDF \eqref{eq:20}, respectively.

	\subsubsection{Problem Formulation}
	
	Based on the above PCRB-based target location distribution exploitation, we aim to design the radar transmit beampattern to minimize the PCRB, thereby improving the radar DoA estimation performance. 
	Specifically, the PCRB oriented transmit beampattern design problem is formulated by minimizing the PCRB subject to the PAPR and transmit power constraints, which is given by
	\begin{subequations}
		\begin{numcases}{\mathcal{P}^{1-1}_{{\rm PCRB},{\bf X}}}
		\mathop {\min }\limits_{\bf{X}} \;\;{ {{\rm{PCRB}}} _\theta }\left( {\bf{X}} \right) \label{eq:P1-1-a}\\
		\;{\rm{s}}.{\rm{t}}. \; \left\| {\bf{X}} \right\|_F^2 = P \label{eq:P1-1-b}\\
		\qquad \; \frac{{{M_t}L}}{{ P}}\mathop {\max }\limits_{m,l} {\left| {{\bf{X}}\left( {m,l} \right)} \right|^2}  \le \kappa,  \forall m,l,  \label{eq:P1-1-c} 
		\end{numcases}\label{eq:P1-1}%
	\end{subequations}
	where $\kappa$ represents the PAPR requirement.
	
	Before proceeding, we make the following remarks.
	\begin{remark}[Is minimizing PCRB effective?]\label{remark:3}
		It is worth mentioning that the derived PCRB \eqref{eq:15} is just a theoretical lower bound for MSE of estimation of the target angular location $\theta$.
		Although the problem \eqref{eq:P1-1} is to minimize the PCRB, whether a lower bound can be reached in practice depends on various factors.
		This raises a critical question: \textit{``Is minimizing the PCRB effective?"} Or \textit{``Is there any other method to effectively exploit target location distribution?"}
	\end{remark}
	
	\vspace{-1em}
	\subsection{PSBP-based Target Location Distribution Exploitation}
	
	It is widely acknowledged that transmit beampattern should be focused on the potential direction of the target to enhance radar detection and estimation performance. 
	This idea motivates us to address \textit{Remark \ref{remark:3}} from the beampattern design perspective and explore another way to exploit target location distribution in the following.
	
	\subsubsection{PSBP Metric}
	The radar transmit power towards angle $\theta$ is defined as $ \mathcal{E}(\theta) = \| {{\bf{a}}_t^H( {{\theta _p}} ){\bf{X}}} \|_F^2 $.
	Note that $ \mathcal{E}(\theta) = \| {{\bf{a}}_t^H( {{\theta}} ){\bf{X}}} \|_F^2 $ is highly dependent on $\theta$ without exploiting the target location distribution. 
	Therefore, a novel radar beampattern metric, called PSBP, is proposed to characterize the radar transmit power towards $\theta_p$ scaled by the PDF $f(\theta_p)$.
	Specifically, the PSBP at the $p$-th point in the possible target angular location region is defined as
	\vspace{-1em}
	\begin{equation}\label{eq:22}
	{\rm{PSBP}}_{{\theta _p}}^w \left( {\bf{X}} \right) = w\left( {f\left( {{\theta _p}} \right)} \right)\left\| {{\bf{a}}_t^H\left( {{\theta _p}} \right){\bf{X}}} \right\|_F^2,\;{\theta _p} \in {\bf{\Theta }},
	\end{equation}
	where $w\left( {f\left( {{\theta _p}} \right)} \right)$ denotes the scaling weight function of the angular location PDF ${f\left( {{\theta _p}} \right)}$ at angle $\theta_p$.
	The specific form of $w\left( {f\left( {{\theta _p}} \right)} \right)$ is determined by the requirements of transmit beampattern optimization.

	\subsubsection{Problem Formulation}
	Based on the newly proposed PSBP, we present the following two PSBP oriented transmit beampattern problems for different design requirements.
	
	\textbf{Fair PSBP Design.}
	Based on the target location distribution model, the target of interest occurs at different possible locations with different probabilities. 
	To ensure the beampattern is fair towards all possible locations, we propose a fair PSBP design problem, where the minimal PSBP is maximized. 
	Therefore, the design problem can be formulated as follows
	\begin{subequations}
		\begin{numcases}{\mathcal{P}^{2-1}_{{\rm PSBP},{\bf X}}}
		\mathop {\max }\limits_{\bf{X}} \;\;\mathop {\min }\limits_{{\theta _p} \in {\bf{\Theta }}} \;{\rm{PSBP}}_{{\theta _p}}^w\left( {\bf{X}} \right)\label{eq:P2-1-a}\\
		\;{\rm{s}}.{\rm{t}}.\;\; \left\| {\bf{X}} \right\|_F^2 = P \label{eq:P2-1-b}\\
		\qquad \;\; \frac{{{M_t}L}}{{ P}}\mathop {\max }\limits_{m,l} {\left| {{\bf{X}}\left( {m,l} \right)} \right|^2}  \le \kappa,  \forall m,l, \label{eq:P2-1-c} 
		\end{numcases}\label{eq:P2-1}%
	\end{subequations}
	where the scaling weight function is selected as  $w(f(\theta_p)) = \frac{1}{f(\theta_p)}$, leading to
	${{\rm{PSBP}}_{{\theta _p}}^w}( {\bf{X}} ) =  \frac{1}{{f( {{\theta _p}})}} \| {{\bf{a}}_t^H( {{\theta _p}} ){\bf{X}}} \|_F^2$.

	\textbf{Integrated PSBP Design.}
	In addition to maintaining fairness towards all possible locations, it is straightforward to transmit more power towards directions with higher probabilities. 
	Thus, we propose an integrated PSBP design problem, where the integrated PSBP is maximized. 
	Therefore, the design problem can be formulated as follows:
	\begin{subequations}
		\begin{numcases}{\mathcal{P}^{3-1}_{{\rm PSBP},{\bf X}}}
		\mathop {\max }\limits_{\bf{X}} \;\;\textstyle\sum\limits_{{\theta _p} \in {\bf{\Theta }}} {{\rm{PSB}}{{\rm{P}}_{{\theta _p}}^w}\left( {\bf{X}} \right)}  \label{eq:P3-1-a}\\
		\;{\rm{s}}.{\rm{t}}.\quad \left\| {\bf{X}} \right\|_F^2 = P \label{eq:P3-1-b}\\
		\qquad \quad \frac{{{M_t}L}}{{ P}}\mathop {\max }\limits_{m,l} {\left| {{\bf{X}}\left( {m,l} \right)} \right|^2}  \le \kappa,  \forall m,l,  \label{eq:P3-1-c} 
		\end{numcases}\label{eq:P3-1}%
	\end{subequations}
	where the scaling weight function is selected as  $w(f(\theta_p)) = {f(\theta_p)}$, leading to
	${\rm{PSB}}{{\rm{P}}_{{\theta _p}}^w}( {\bf{X}} ) =  {{f( {{\theta _p}})}} \| {{\bf{a}}_t^H( {{\theta _p}} ){\bf{X}}} \|_F^2$.
	
	{Note that the radar transmit beampattern design problems \eqref{eq:P1-1}, \eqref{eq:P2-1} and \eqref{eq:P3-1} are all non-convex, which is hard to settle by the existing methods.}
	Before moving to solutions to design problems, we make the following remarks.
	\begin{remark}[Non-recursive beampattern design]
		Compared with existing recursive PCRB-based transmit beampattern designs, the proposed non-recursive approach offers the following advantages:
		\textit{i)} Direct and global optimization: It performs a one-shot optimization that directly incorporates the entire prior distribution, avoiding posterior feedback or a running point estimate in recursive designs and significantly reducing computational cost.
		\textit{ii)} No error accumulation: Recursive schemes rely on previous posterior updates, which may propagate estimation errors across iterations. The non-recursive method inherently avoids this issue and ensures more stable performance.
		\textit{iii)} Comprehensive exploitation of prior information: Instead of exploiting posterior information gradually and approximately, the proposed framework fully leverages the complete prior knowledge, leading to improved robustness under uncertainty.
	\end{remark}
	\vspace{-1em}
	
	\begin{remark}[Comparison between PCRB and PSBP]
		\textit{i)} Compared with PSBP, PCRB has a more specific meaning, denoting the lower bound of radar estimation performance. 
		However, although many studies show that designing radar transmit beampattern with proper direction can improve radar estimation performance, the mathematical relationship between the PSBP and estimation performance is still unclear.
		\textit{ii)} PCRB requires pre-calculating many distribution-determined parameters, ${{{\bf{\Xi }}_1}}$, ${{{\bf{\Xi }}_2}}$, ${{{\bf{\Xi }}_3}}$, ${\Lambda }$, leading to higher computational complexity. 
		In contrast, the newly proposed PSBP directly incorporates the prior information through a PDF-scaled transmit power function, enabling a more compact and explicit formulation and eliminating the need for computationally intensive pre-calculations (e.g., high-dimensional integration) to evaluate the prior-dependent FIM.
	\end{remark}
	
	\vspace{-1.5em}
	\section{Solution to PCRB Oriented transmit beampattern Design}
	
	In this section, we first simplify the PCRB oriented transmit beampattern design problem, then derive a solution, and finally conclude with a summary of the proposed method.
	
	\vspace{-1em}
	\subsection{Problem Reformulation}
	
	\subsubsection{Step 1} \textit{Simplification to $\mathrm{PCRB}_\theta(\mathbf{X})$.}
	Since the objective function ${\rm{PCR}}{{\rm{B}}_\theta }\left( {\bf{X}} \right)$ in \eqref{eq:P1-1} is a complicated fractional function with the inverse operation, we present the following proposition to simplify it into a more tractable form.
	\begin{prop}\label{prop:2}
		We introduce an upper bound of PCRB, $\overline {{\rm{PCRB}}}_{\theta} \left( \bf X \right)$ as
		\begin{equation}\label{eq:24}
		\begin{aligned}
		{\rm{PCR}}{{\rm{B}}_\theta }\left( {\bf{X}} \right) & < {\left[ {\Lambda  + \frac{{2{{\left| \varsigma  \right|}^2}}}{{\sigma _{\rm{r}}^2}}{\rm{Tr}}\left\{ {{{\bf{X}}^H}{{\bf{\Xi }}_0}{\bf{X}}} \right\}} \right]^{ - 1}}  \\
		& \buildrel \Delta \over = {\overline {{\rm{PCRB}}} _\theta }\left( {\bf{X}} \right) ,
		\end{aligned}
		\end{equation}
		where ${{\bf{\Xi }}_0} = \int_{ - \pi /2}^{\pi /2} {f\left( \theta  \right)\left\| {{{{\bf{\dot a}}}_r}\left( \theta  \right)} \right\|_F^2{{\bf{a}}_t}\left( \theta  \right){\bf{a}}_t^H\left( \theta  \right)d\theta }$.
	\end{prop}
	\begin{IEEEproof}
		Please refer to Appendix \ref{app:B}.
	\end{IEEEproof}
	According to \textit{Proposition \ref{prop:2}} and noting that $\Lambda$ is irrelevant to $\mathbf{X}$, the $\mathcal{P}^{1-1}_{{\rm PCRB},{\bf X}}$ can be simplified as
	\begin{subequations}
		\begin{numcases}{\mathcal{P}^{1-2}_{{\rm PCRB},{\bf X}}}
		\mathop {\min }\limits_{{\bf{X}}} \;\; - {\rm{Tr}}\left\{ {{{\bf{X}}^H}{{\bf{\Xi }}_0}{\bf{X}}} \right\} \\
		\;{\rm{s}}.{\rm{t}}.\;\; \left\| {\bf{X}} \right\|_F^2 = P \\
		\qquad \;\;{\left| {{\bf{X}}\left({m,l}\right) } \right|^2} \le \frac{{\kappa P}}{{{M_t}L}},\forall m,l  .
		\end{numcases}
	\end{subequations}
	
	\subsubsection{Step 2} \textit{Application of ADMM Framework to ${\mathcal{P}^{1-2}_{{\rm PCRB},{\bf X}}}$.}
	To derive an efficient solution to ${\mathcal{P}^{1-2}_{{\rm PCRB},{\bf X}}}$, we introduce an auxiliary variable $\mathbf{U}$ to decouple the objective and constraints, leading to the following problem.
	\begin{subequations}
		\begin{numcases}{\mathcal{P}^{1-3}_{{\rm PCRB},{\bf X}}}
		\mathop {\min }\limits_{{\bf{X}},{\bf U}} \;\; - {\rm{Tr}}\left\{ {{{\bf{X}}^H}{{\bf{\Xi }}_0}{\bf{X}}} \right\} \label{eq:P1-2-a}\\
		\;{\rm{s}}.{\rm{t}}.\quad \left\| {\bf{X}} \right\|_F^2 = P \label{eq:P1-2-b}\\
		\qquad \quad{\left| {{\bf{U}}\left({m,l}\right) } \right|^2} \le \frac{{\kappa P}}{{{M_t}L}},\forall m,l \label{eq:P1-2-c}\\
		\qquad \quad{\bf{X}} = {\bf{U}} . \label{eq:P1-2-d}
		\end{numcases}\label{eq:P1-2}%
	\end{subequations}
	By penalizing the equality constraint \eqref{eq:P1-2-d} into the objective function, we equivalently rewrite the problem \eqref{eq:P1-2} into an augmented Lagrangian (AL) minimization problem as follows
	\begin{subequations}
		\begin{numcases}{\mathcal{P}^{1-4}_{{\rm PCRB},{\bf X}}}
		\mathop {\min }\limits_{{{\bf{X}},{\bf{U}},{\bf{D}}_1}} \;\;{\cal L}{_1}\left( {{\bf{X}},{\bf{U}},{\bf{D}}_1} \right) \label{eq:P1-3-a}\\
		\quad{\rm{s}}.{\rm{t}}.\quad \eqref{eq:P1-2-b} \text{ and } \eqref{eq:P1-2-c},
		\end{numcases}\label{eq:P1-3}%
	\end{subequations}
	The associated AL function is given by
	\begin{equation}\label{eq:L1}
	{\cal L}{_1}( {{\bf{X}},{\bf{U}},{\bf{D}}_1} ) 
	=  - {\rm{Tr}}\{ {{{\bf{X}}^H}{{\bf{\Xi }}_0}{\bf{X}}} \}
	+ \frac{{{\rho _1}}}{2}\| {{\bf{U}} - {\bf{X}} + {{\bf{D}}_1}} \|_F^2 ,\nonumber
	\end{equation}
	where $\rho_1 > 0$ is a penalty parameter and ${\bf D}_1$ is the dual variable.
	Under an ADMM framework, we can update $\{{\bf X},{\bf U},{\bf D}_1\}$ by taking the following iterative steps. 
	\begin{subequations}
		\begin{align}
		& {{\bf{U}}^{t + 1}}: = \arg \;\min_{\bf U} \;{{\mathcal L}_1}\left( {{{\bf{X}}^{t}},{\bf{U}},{\bf{D}}_1^t} \right), \quad \mathrm{s.t.} \; \eqref{eq:P1-2-c} , \label{eq:26-a}\\
		&{{\bf{X}}^{t + 1}}: = \arg \;\min_{\bf X} \;{{\mathcal L}_1}\left( {{\bf{X}},{{\bf{U}}^{t+1}},{\bf{D}}_1^t} \right), \;  \mathrm{s.t.} \; \eqref{eq:P1-2-b},\label{eq:26-b} \\
		& {\bf{D}}_1^{t + 1}: = {\bf{D}}_1^t + {{\bf{U}}^{t + 1}} - {{\bf{X}}^{t + 1}}. \label{eq:26-c} 
		\end{align}
	\end{subequations}%
	where $t$ is iteration number, $( \cdot )^t$ is the last point of $( \cdot )^{t+1}$.

	\subsection{Solutions to the Subproblems \eqref{eq:26-a}-\eqref{eq:26-b}}
	
	\subsubsection{Optimization of $\bf U$}
	Given $\bf X$ and ${\bf D}_1$, $\bf U$ can be updated by solving 
	\begin{equation}\label{neq:31}
	\mathop {\min }\limits_{ {\bf{U}} } \; \left\| \mathbf{U} - \mathbf{X} + \mathbf{D}_1 \right\|_F^2, \; {\rm{s}}.{\rm{t}}.\;{\left| {{\bf{U}}\left( {m,l} \right)} \right|^2} \le \frac{{\kappa P}}{{{M_t}L}},\forall m,l .
	\end{equation}
	Problem \eqref{neq:31} can be decoupled into the following $ML$ consensus problem.
	\begin{equation}\label{eq:31}
	\begin{aligned}
	&\mathop {\min }\limits_{ {\bf{U}}(m,l) } \;{\left| {{\bf{U}}\left( {m,l} \right) - {\bf{X}}\left( {m,l} \right) + {{\bf{D}}_1}\left( {m,l} \right)} \right|^2}, \\
	&\quad{\rm{s}}.{\rm{t}}.\quad{\left| {{\bf{U}}\left( {m,l} \right)} \right|^2} \le \frac{{\kappa P}}{{{M_t}L}},\forall m,l ,
	\end{aligned}
	\end{equation}
	whose optimal solution can be directly calculated by
	\begin{equation}
	{\bf{U}}^{t+1}\left( {m,l} \right) = \sqrt {\frac{{\kappa P}}{{{M_t}L}}} \frac{{{\bf{X}}\left( {m,l} \right) - {{\bf{D}}_1}\left( {m,l} \right)}}{{\left| {{\bf{X}}\left( {m,l} \right) - {{\bf{D}}_1}\left( {m,l} \right)} \right|}},\;\forall m,l.
	\end{equation}

	\subsubsection{Optimization of $\bf X$}
	Given $\bf U$ and ${\bf D}_1$, $\bf X$ can be updated by solving
	\begin{subequations}
		\begin{align}
		&\mathop {\min }\limits_{\bf{X}} \; - {\rm{Tr}}\left\{ {{{\bf{X}}^H}{{\bf{\Xi }}_0}{\bf{X}}} \right\} + \frac{{{\rho _1}}}{2}\left\| {{\bf{U}} - {\bf{X}} + {{\bf{D}}_1}} \right\|_F^2, \label{eq:27-a}\\
		&\;\;{\rm{s}}.{\rm{t}}.\;\;\left\| {\bf{X}} \right\|_F^2 = P . \label{eq:27-b}
		\end{align}\label{eq:27}%
	\end{subequations}
	By utilizing the Karush-Kuhn-Tucker (KKT) conditions, the closed form solution to $\bf X$ can be calculated by
	\begin{equation}\label{eq:28}
	{\bf{X}}\left( {{\mu _1}} \right) = {\left( {{{\bf{P}}_1} + 2{\mu _1}{{\bf{I}}_{{M_t}}}} \right)^{ - 1}}{{\bf{Q}}_1} ,
	\end{equation}
	where $\mu_1$ is the Lagrange multiplier, ${{\bf{P}}_1} = {\rho _1}{{\bf{I}}_{{M_t}}} - \left( {{{\bf{\Xi }}_0} + {\bf{\Xi }}_0^H} \right)$ and ${{\bf{Q}}_1} = {\rho _1}\left( {{\bf{U}} + {{\bf{D}}_1}} \right)$.
	
	To find the optimal solution $\mu_1$, we define the eigen-decomposition of ${{\bf{P}}_1}$ as ${{\bf{P}}_1} = {{\bf{G}}_1}{{\bf{\Sigma }}_1}{\bf{G}}_1^H$.
	Then, \eqref{eq:28} can be rewritten as
	\begin{equation}\label{eq:29}
	{\bf{X}}\left( {{\mu _1}} \right) = {{\bf{G}}_1}{\left( {{{\bf{\Sigma }}_1} + 2{\mu _1}{{\bf{I}}_{{M_t}}}} \right)^{ - 1}}{\bf{G}}_1^H{{\bf{Q}}_1} .
	\end{equation}
	Substituting \eqref{eq:29} into the constraint \eqref{eq:27-b}, we can obtain
	\begin{equation}\label{eq:30}
	\left\| {\bf{X}}\left( {{\mu _1}} \right) \right\|_F^2 = \textstyle\sum\limits_{m = 1}^{{M_t}} {\frac{{{{\bf{\Psi }}_1}\left( {m,m} \right)}}{{{{\left( {{{\bf{\Sigma }}_1}\left({m,m} \right) + 2{\mu _1}} \right)}^2}}}}  = P ,
	\end{equation}
	where ${{\bf{\Psi }}_1} = {\bf{G}}_1^H{{\bf{Q}}_1}{\bf{Q}}_1^H{{\bf{G}}_1}$.
	Based on \eqref{eq:30},  {$\sum\nolimits_{m = 1}^{{M_t}} {\frac{{{{\bf{\Psi }}_1}\left( {m,m} \right)}}{{{{\left( {{{\bf{\Sigma }}_1}\left({m,m} \right) + 2{\mu _1}} \right)}^2}}}}$ is strictly decreasing w.r.t. $\mu_1 > 0$. Therefore,} the optimal solution $\mu_1^{t+1}$ can be found using Newton's or bisection method.
	Finally, the optimal ${\bf X}^{t+1}$ can be calculated by substituting the optimal $\mu_1^{t+1}$ into \eqref{eq:29}.
	
	\vspace{-1em}
	\subsection{Complexity and Convergence Performance Analysis}

	Finally, we analyze the complexity of the proposed PCRB oriented transmit beampattern design method and present the convergence performance of the proposed algorithm.
	
	\subsubsection{Complexity Analysis}
	The complexity of the proposed PCRB oriented transmit beampattern design includes two parts as follows.
	
	$\bullet$~\textbf{Pre-calculation of distribution parameters:}
	Computing ${{\bf{\Xi }}_0},{{\bf{\Xi }}_1}, {{\bf{\Xi }}_2}, {{\bf{\Xi }}_3}$ with integration operation need the complexity of ${\cal O}\left( {M_t^2{N_{{\rm{int}}}}} \right)$ and computing ${\Lambda}$ needs the complexity of ${\cal O}\left( {K^2{N_{{\rm{int}}}}} \right)$, where ${\cal O}\left( {{N_{{\rm{int}}}}} \right)$ represents the complexity of integrating a function.
	
	$\bullet$~\textbf{Updating in beampattern design algorithm:}
	Updating $\bf X$ needs a complexity of ${\mathcal O}(M_t^3)$ and updating $\bf U$ needs a complexity of ${\mathcal O}(M_tL)$.
	The complexity of the proposed PCRB oriented algorithm is ${\cal O}\left( N_{\rm ite}\left(M_t^3 + M_tL\right)\right)$, where $N_{\rm ite}$ is the iteration number in the ADMM framework.
	
	To sum up, the total complexity of the proposed PCRB oriented transmit beampattern design method is ${\cal O}\left( N_{\rm ite}\left(M_t^3 + M_tL\right)  + {{N_{{\rm{int}}}}\left(K^2+M_t^2\right)} \right)$.
	
	\subsubsection{Convergence Performance}\label{sec:IV-C-2}
	
	Convergence of the proposed algorithm is characterized by the following proposition.
	\begin{prop}\label{prop:3}
		For ${\rho _1} > \sqrt 3 {\| { {{{\bf{\Xi }}_0} + {\bf{\Xi }}_0^H} }\|_F}$, the sequence $\{{\bf X}^{t}, {\bf U}^{t}, {\bf D}^{t}\}$ generated by the ADMM-based transmit beampattern design algorithm via PCRB oriented method has the following properties
		\begin{enumerate}
			\item The augmented Lagrange function ${\cal L}{_1}$ is decent during $\{{{\bf{X}},{\bf{U}},{\bf{D}}_1}\}$ update;
			\item The augmented Lagrange function ${\cal L}{_1}({{\bf{X}},{\bf{U}},{\bf{D}}_1})$ is lower bounded for all $t$ and converges as $t \to  + \infty $;
			\item  The residual error ${\lim _{t \to \infty }}\left\| {{{\bf{U}}^t} - {{\bf{X}}^t}} \right\|_F^2 = 0$.
		\end{enumerate}
	\end{prop}
	\begin{IEEEproof}
		Please refer to Appendix \ref{app:C}.
	\end{IEEEproof}
	
	\vspace{-1em}
	\section{Solution to PSBP Oriented transmit beampattern Design}
	
	In this section, we first transform the fair PSBP transmit beampattern design problem \eqref{eq:P2-1} into a more tractable form, and derive a corresponding solution.
	Then, we extend the solution to the integrated PSBP transmit beampattern design \eqref{eq:P3-1}.
	Finally, a summary of the proposed algorithms for both problems is given.
	
	\vspace{-1em}
	\subsection{Problem Reformulation of \eqref{eq:P2-1}}
	
	\subsubsection{Step 1} \textit{Reformulation of \eqref{eq:P2-1}.}
	Since the problem \eqref{eq:P2-1} is a complicated max-min problem, we first equivalently transform it to a more tractable form by introducing an auxiliary variable $\eta$ as
	\begin{subequations}
		\begin{numcases}{\mathcal{P}^{2-2}_{{\rm PSBP},{\bf X}}}
		\mathop {\min }\limits_{\bf{X}} - \eta \label{eq:P2-2-a}\\
		\;{\rm{s}}.{\rm{t}}. \left\| {\bf{X}} \right\|_F^2 = P \label{eq:P2-2-b}\\
		\quad{\left| {{\bf{X}}\left( {m,l} \right)} \right|^2} \le \frac{{\kappa P}}{{{M_t}L}},\forall m,l \label{eq:P2-2-c} \\
		\quad\frac{1}{{f\left( {{\theta _p}} \right)}}\left\| {{\bf{a}}_t^H\left( {{\theta _p}} \right){\bf{X}}} \right\|_F^2 \ge \eta ,\forall {\theta _p} \in {\bf{\Theta }}, \label{eq:P2-2-d} \\
		{  \quad\eta > 0,} \label{eq:P2-2-e}
		\end{numcases}\label{eq:P2-2}%
	\end{subequations}
	
	\subsubsection{Step 2} \textit{Application of ADMM Framework to ${\mathcal{P}^{2-2}_{{\rm PSBP},{\bf X}}}$.}
	To derive an efficient solution to ${\mathcal{P}^{2-2}_{{\rm PSBP},{\bf X}}}$, we propose introducing several auxiliary variables $\bf T$ and $\{{\bf{g}}_p\}$ to decouple the constraints, leading to the following problem.
	\begin{subequations}
		\begin{numcases}{\mathcal{P}^{2-3}_{{\rm PSBP},{\bf X}}}
		\mathop {\min }\limits_{{\bf{X}},{\bf T},\{{\bf g}_p\}} \;\; - \eta \label{eq:P2-3-a}\\
		\quad{\rm{s}}.{\rm{t}}.\quad \left\| {\bf{X}} \right\|_F^2 = P \label{eq:P2-3-b}\\
		\qquad \quad\;\;{\left| {{\bf{T}}\left( {m,l} \right)} \right|^2} \le \frac{{\kappa P}}{{{M_t}L}},\forall m,l \label{eq:P2-3-c} \\
		\qquad\quad\;\;\frac{1}{{f\left( {{\theta _p}} \right)}}\left\| {{{{\bf{g}}_p}}} \right\|_F^2 \ge \eta ,\forall {\theta _p} \in {\bf{\Theta }} \label{eq:P2-3-d} \\
		{  \qquad\quad\;\;\eta > 0.} \label{eq:P2-3-e}\\
		\qquad\quad\;\;{\bf{T}} = {\bf{X}} \label{eq:P2-3-f} \\
		\qquad\quad\;\;{{\bf{g}}_p} = {{\bf{X}}^H}{{\bf{a}}_t}\left( {{\theta _p}} \right),\forall {\theta _p} \in {\bf{\Theta }}.\label{eq:P2-3-g}
		\end{numcases}\label{eq:P2-3}%
	\end{subequations}
	By penalizing the equality constraints \eqref{eq:P2-3-e} and \eqref{eq:P2-3-f} into the objective function, the problem \eqref{eq:P2-3} can be equivalently rewritten into an AL minimization problem:
	\begin{subequations}
		\begin{numcases}{\mathcal{P}^{2-4}_{{\rm PSBP},{\bf X}}}
		\mathop {\min }\limits_{\scriptstyle\eta ,{\bf{X}},{\bf{T}},\{ {{\bf{g}}_p}\} ,\hfill\atop
			\scriptstyle\;\;{{\bf{D}}_2},\{ {{\bm{\beta }}_p}\} \hfill} {{\cal L}_2}\left({\eta}, {\bf{X}},{\bf{T}},{\{ {{\bf{g}}_p}\}}, {{\bf D}_2}, \{{\bm{\beta }}_p\}\right) \label{eq:P2-4-a}\\
		\quad\;\;{\rm{s}}.{\rm{t}}.\quad \eqref{eq:P2-3-b}- {\eqref{eq:P2-3-e}}.
		\end{numcases}\label{eq:P2-4}%
	\end{subequations}
	The associated AL function via penalizing the equality constraints \eqref{eq:P2-3-e} and \eqref{eq:P2-3-f} is given by
	\begin{equation}
	\begin{aligned}
	&{{\cal L}_2}\left({\eta}, {\bf{X}},{\bf{T}},{\{ {{\bf{g}}_p}\}}, {{\bf D}_2}, \{{\bm{\beta }}_p\}\right) \! = \!- \eta  + \frac{{{\rho _2}}}{2}\left\| {{\bf{T}} - {\bf{X}} + {{\bf{D}}_2}} \right\|_F^2 \\
	&\qquad\qquad\qquad\qquad+ \frac{{{\rho _3}}}{2}\textstyle\sum\limits_{{\theta _p} \in {\bf{\Theta }}} {\left\| {{{\bf{g}}_p} - {{\bf{X}}^H}{{\bf{a}}_t}\left( {{\theta _p}} \right) + {{\bm{\beta }}_p}} \right\|_F^2} ,
	\end{aligned}\nonumber
	\end{equation}
	where $\rho_2,\rho_3 >0$ are penalty parameters, ${{{\bf{D}}_2}}$ and ${\{ {{\bm{\beta }}_p}\} }$ are dual variables.
	Similarly, under an ADMM framework, we can update $\left\{ {{\bf{X}},{\bf{T}},\{ {{\bf{g}}_p}\} } \right\}$ by taking the following iterative steps.
	\begin{subequations}
		\begin{align}
		&{{\bf{T}}^{t + 1}}: = \arg \;\min_{\bf T} \;{{\cal L}_2}\left({\eta}^t, {\bf{X}}^{t},{\bf{T}},{\{ {{\bf{g}}_p^t}\}}, {{\bf D}_2^t}, \{{\bm{\beta }}_p^t\}\right),\nonumber\\
		&\qquad\qquad\qquad {\rm{s.t.}} \;\eqref{eq:P2-3-c},\label{eq:42-a}\\
		&{{\bf{X}}^{t + 1}}: = \arg \;\min_{\bf X} \;{{\cal L}_2}\left({\eta}^t, {\bf{X}},{\bf{T}}^{t+1},{\{ {{\bf{g}}_p^t}\}}, {{\bf D}_2^t}, \{{\bm{\beta }}_p^t\}\right), \nonumber
		\\&\qquad\qquad\qquad {\rm{s.t.}}\; \eqref{eq:P2-3-b},\label{eq:42-b}\\
		& {{\eta ^{t + 1}},\{ {\bf{g}}_p^{t + 1}\} }: =\nonumber\\
		&\qquad\arg \mathop {\min }\limits_{\eta ,\{ {{\bf{g}}_p}\} } {{\cal L}_2}\left({\eta},{\bf{X}}^{t+1},{\bf{T}}^{t+1},{\{ {{\bf{g}}_p}\}}, {{\bf D}_2^t}, \{{\bm{\beta }}_p^t\}\right),\nonumber\\
		&\qquad \qquad\;{\rm{s.t.}}\;\eqref{eq:P2-3-d},\eqref{eq:P2-3-e}\label{eq:42-c}\\
		& {\bf{D}}_2^{t + 1}: = {\bf{D}}_2^t + {{\bf{T}}^{t + 1}} - {{\bf{X}}^{t + 1}}, \label{eq:42-d}\\
		& {\bm{\beta }}_p^{t + 1}: = {\bm{\beta }}_p^t + {\bf{g}}_p^{t + 1} - {\left( {{{\bf{X}}^{t + 1}}} \right)^H}{{\bf{a}}_t}\left( {{\theta _p}} \right),{\theta _p} \in {\bf{\Theta }}, \label{eq:42-e}
		\end{align}
	\end{subequations}%
	Note that the subproblems \eqref{eq:42-a} and \eqref{eq:42-b} can be solved similarly to the subproblems \eqref{eq:26-a} and \eqref{eq:26-b}, respectively.
	To avoid redundancy, we omit the solutions to the subproblems \eqref{eq:42-a} and \eqref{eq:42-b}, and only present the solution to the subproblem \eqref{eq:42-c} as follows.
	
	\vspace{-0.5em}
	\subsection{Solutions to the Subproblem \eqref{eq:42-c}}
	
	\subsubsection{Optimization of $\eta,\{ {{\bf{g}}_p}\}$}
	Given other variables, $\eta,\{ {{\bf{g}}_p}\}$ can be updated by solving
	\begin{subequations}
		\begin{align}
		&\mathop {\min }\limits_{\eta,\{ {{\bf{g}}_p}\}} \;\;\; - \eta  + \frac{{{\rho _3}}}{2}\textstyle\sum\limits_{{\theta _p} \in {\bf{\Theta }}} {\left\| {{{\bf{g}}_p} - {{\bf{h}}_p}} \right\|_F^2}, \label{eq:44-a} \\
		&\;\;{\rm{s}}.{\rm{t}}.\quad \frac{1}{{f\left( {{\theta _p}} \right)}}\left\| {{{\bf{g}}_p}} \right\|_F^2 \ge \eta ,\forall {\theta _p} \in {\bf{\Theta }}, \label{eq:44-b}\\
		&{  \qquad\quad\eta > 0,\label{eq:44-c}} 
		\end{align}\label{eq:44}%
	\end{subequations}
	where we define ${{\bf{h}}_p} = {{\bf{X}}^H}{{\bf{a}}_t}\left( {{\theta _p}} \right) - {{\bm{\beta }}_p}$.
	To find the optimal solution to $\{ {{\bf{g}}_p}\} ,\eta$, we take the following steps:
	
	$\bullet$~{Step 1:} $\eta$ is fixed, and $\{{{\bf{g}}_p}\}$ is updated by
	\begin{equation}\label{eq:45}
	\mathop {\min }\limits_{\{ {{\bf{g}}_p}\} } \;\left\| {{{\bf{g}}_p} - {{\bf{h}}_p}} \right\|_F^2
	\;\;{\rm{s}}.{\rm{t}}.\;\left\| {{{\bf{g}}_p}} \right\|_F^2 \ge f\left( {{\theta _p}} \right)\eta ,\forall {\theta _p} \in {\bf{\Theta }},
	\end{equation} 
	whose solution can be calculated as
	\begin{equation}
	{{\bf{g}}_p} = \left\{ \begin{array}{l}
	\qquad {{\bf{h}}_p},\qquad\quad\quad \left\| {{{\bf{h}}_p}} \right\|_F^2 \ge f\left( {{\theta _p}} \right)\eta \\
	\sqrt {f\left( {{\theta _p}} \right)\eta } \frac{{{{\bf{h}}_p}}}{{{{\left\| {{{\bf{h}}_p}} \right\|}_F}}},\;\;{\text{otherwise}}
	\end{array} \right..
	\end{equation}
	
	$\bullet$~{Step 2:} Substituting the updated $\{{{\bf{g}}_p}\}$,  {for $\eta > 0$,} the problem \eqref{eq:44} can be converted into
	\begin{equation}
	\mathop {\min }\limits_\eta  \;f\left( \eta  \right) =  - \eta  + \frac{{{\rho _3}}}{2}\textstyle\sum\limits_{{\theta _p} \in {\bf{\Theta }}} {{\varpi _p}{\left( {\sqrt {f\left( {{\theta _p}} \right)} {\eta ^{1/2}} - {{\left\| {{{\bf{h}}_p}} \right\|}_F}} \right)^2}} ,\nonumber
	\end{equation}
	where ${\varpi _p} = \left\{ \begin{array}{l}
	0,\; \left\| {{{\bf{h}}_p}} \right\|_F^2 \ge f\left( {{\theta _p}} \right)\eta \\
	1,\;{\text{otherwise}}
	\end{array} \right.$.
	By taking the first and second order derivatives of $f\left( \eta  \right)$, we can obtain
	\begin{equation}
	\begin{aligned}
	&\frac{{\partial f\left( \eta  \right)}}{{\partial \eta }} \!=\!  - 1 \!+\! \frac{{{\rho _3}}}{2}\!\!\textstyle\sum\limits_{{\theta _p} \in {\bf{\Theta }}}\!\!\!\!{\sqrt {f\left( {{\theta _p}} \right)} {\varpi _p}( {\sqrt {f\left( {{\theta _p}} \right)}  \!-\! {{\left\| {{{\bf{h}}_p}} \right\|}_F}{\eta ^{ - 1/2}}} )},\\
	&\frac{{{\partial ^2}f\left( \eta  \right)}}{{{\partial ^2}\eta }} = \frac{{{\rho _3}}}{4}\textstyle\sum\limits_{{\theta _p} \in {\bf{\Theta }}} {\sqrt {f\left( {{\theta _p}} \right)} {\varpi _p}{{\left\| {{{\bf{h}}_p}} \right\|}_F}{\eta ^{ - 3/2}}} .\nonumber
	\end{aligned}
	\end{equation}
	Obviously, $\frac{{{\partial ^2}f\left( \eta  \right)}}{{{\partial ^2}\eta }} \ge 0$ and then $\frac{{\partial f\left( \eta  \right)}}{{\partial \eta }}$ is monotonically increasing.
	Applying the first-order optimality condition, we can obtain the optimal $\eta^{t+1}$ by solving $\frac{{\partial f\left( \eta  \right)}}{{\partial \eta }} = 0$ with Newton's or bisection method.
	
	\vspace{-1em}
	\subsection{Extension to Solution to the Integrated PSBP transmit beampattern Design Problem \eqref{eq:P3-1}}
	
	To derive an efficient solution to ${\mathcal{P}^{3-1}_{{\rm PSBP},{\bf X}}}$, we introduce an auxiliary variable $\bf V$ to decouple the objective and constraints, leading to the following problem.
	\begin{subequations}
		\begin{numcases}{\mathcal{P}^{3-2}_{{\rm PSBP},{\bf X}}}
		\mathop {\min }\limits_{{\bf{X}},{\bf V}} \;\; - \textstyle\sum\limits_{{\theta _p} \in {\bf{\Theta }}} {f\left( {{\theta _p}} \right)\left\| {{\bf{a}}_t^H\left( {{\theta _p}} \right){\bf{X}}} \right\|_F^2} \label{eq:P3-2-a}\\
		\;{\rm{s}}.{\rm{t}}.\quad \left\| {\bf{X}} \right\|_F^2 = P \label{eq:P3-2-b}\\
		\qquad \quad{\left| {{\bf{V}}\left( {m,l} \right)} \right|^2} \le \frac{{\kappa P}}{{{M_t}L}},\forall m,l \label{eq:P3-2-c}\\
		\qquad \quad{\bf{X}} = {\bf{V}}.  \label{eq:P3-2-d}
		\end{numcases}\label{eq:P3-2}%
	\end{subequations}
	By penalizing the equality constraint \eqref{eq:P3-2-d} into the objective function, we convert the problem \eqref{eq:P3-2} into an equivalent minimization problem as follows.
	\begin{subequations}
		\begin{numcases}{\mathcal{P}^{3-3}_{{\rm PSBP},{\bf X}}}
		\mathop {\min }\limits_{{\bf{X}},{\bf V}} \;\; {{\cal L}_3}\left( {{\bf{X}},{\bf{V}},{{\bf{D}}_3}} \right)\\
		\;{\rm{s}}.{\rm{t}}.\quad \eqref{eq:P3-2-b} \text{ and } \eqref{eq:P3-2-c},
		\end{numcases}\label{eq:P3-3}%
	\end{subequations}
	whose associated AL function via penalizing \eqref{eq:P3-2-d} is 
	\begin{equation}
	\begin{aligned}
	{{\cal L}_3}\left( {{\bf{X}},{\bf{V}},{{\bf{D}}_3}} \right) =&  - \textstyle\sum\limits_{{\theta _p} \in {\bf{\Theta }}} {f\left( {{\theta _p}} \right)\left\| {{\bf{a}}_t^H\left( {{\theta _p}} \right){\bf{X}}} \right\|_F^2}  \\
	&+ \frac{{{\rho _4}}}{2}\left\| {{\bf{V}} - {\bf{X}} + {{\bf{D}}_3}} \right\|_F^2,
	\end{aligned}\nonumber
	\end{equation}
	where $\rho_3>0$ is the penalty parameter and { ${\bf D}_3$} is the dual variable.
	Under an ADMM framework, $\left\{ {{\bf{X}},{\bf{V}},{{\bf{D}}_3}}\right\}$ can be updated by taking the following steps.
	\begin{subequations}
		\begin{align}
		&{{\bf{V}}^{t + 1}}: = \arg \;\mathop {\min }\limits_{\bf{V}} \;{{\mathcal L}_3}\left( {{{\bf{X}}^{t}},{\bf{V}},{\bf{D}}_3^t} \right),\quad{\rm{s.t.}}\;\eqref{eq:P3-2-c},\label{eq:50-a}\\
		&{{\bf{X}}^{t + 1}}: = \arg \;\mathop {\min }\limits_{\bf{X}} \;{{\mathcal L}_3}\left( {{\bf{X}},{{\bf{V}}^{t+1}},{\bf{D}}_3^t} \right), \;{\rm{s.t.}}\; \eqref{eq:P3-2-b},\label{eq:50-b}\\
		&{\bf{D}}_3^{t + 1}: = {\bf{D}}_3^t + {{\bf{V}}^{t + 1}} - {{\bf{X}}^{t + 1}}.\label{eq:50-c}
		\end{align}\label{eq:50}%
	\end{subequations}%
	Note that the subproblems \eqref{eq:50-a} and \eqref{eq:50-b} can be solved by similar methods of solving \eqref{eq:26-a} and \eqref{eq:26-b}.
	To avoid redundancy, we omit the specific derivations.

	\vspace{-1em}
	\subsection{Complexity and Convergence Performance Analysis}
	
	Finally, we analyze the complexity of the proposed two PSBP oriented transmit beampattern design methods and present the convergence performance of the proposed algorithms.
	
	\subsubsection{Complexity Analysis}
	Compared with the PCRB oriented transmit beampattern design method, the complexity of the PSBP oriented methods only involves the updating in the proposed algorithms as follows.
	
	$\bullet$~\textbf{Fair PSBP design algorithm:}
	Updating $\bf X$ needs a complexity of ${\mathcal O}(M_t^3)$ and updating $\bf T$ needs a complexity of ${\mathcal O}(M_tL)$.
	Solving \eqref{eq:44} needs a complexity of ${\mathcal O}(M_tLD)$, where $D$ denotes the sampling number of the angular range.
	Therefore, the overall complexity of the proposed algorithm is ${\mathcal O}(N_{\rm ite}(M_t^3 + M_tLD))$, where $N_{\rm ite}$ is the iteration number in the ADMM framework.
	
	$\bullet$~\textbf{Integrated PSBP design algorithm:}
	Updating $\bf X$ needs a complexity of ${\mathcal O}(M_t^3)$ and updating $\bf V$ needs a complexity of ${\mathcal O}(M_tL)$.
	Thus, the overall complexity of the proposed algorithm is ${\mathcal O}(N_{\rm ite}(M_t^3 + M_tL))$, where $N_{\rm ite}$ is the iteration number in the ADMM framework.
	
	\subsubsection{Convergence Performance}
	Convergence of the proposed two PSBP oriented transmit beampattern design algorithms can be proved similarly to the subsubsection \ref{sec:IV-C-2}.
	To avoid redundancy, we omit the specific derivations.
	
	\vspace{-1em}
	\section{Numerical Simulations}
	
	In this section, we conduct numerical simulations to evaluate the effectiveness of the proposed transmit beampattern design algorithms and verify the proposed target location distribution exploitation methods. 
	We start by introducing the simulation setup, followed by three different scenarios corresponding to different distribution models.
	
	\vspace{-1em}
	\subsection{Simulation Setup}
	
	\begin{figure*}[!ht]
		\centering
		\subfigure[]{
			\includegraphics[width=0.32\linewidth]{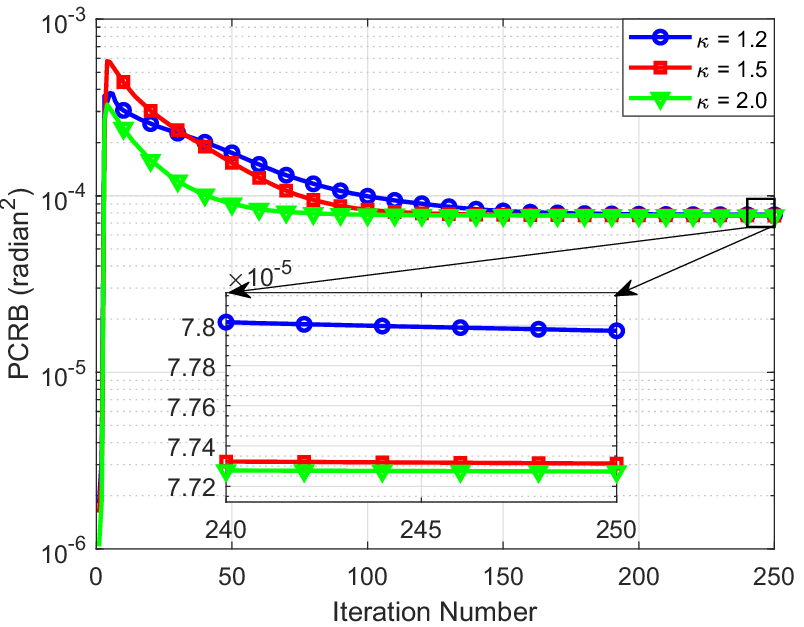} 
			\label{fig:1-1-1}
		}\hspace{-0.5em}
		\subfigure[]{
			\includegraphics[width=0.32\linewidth]{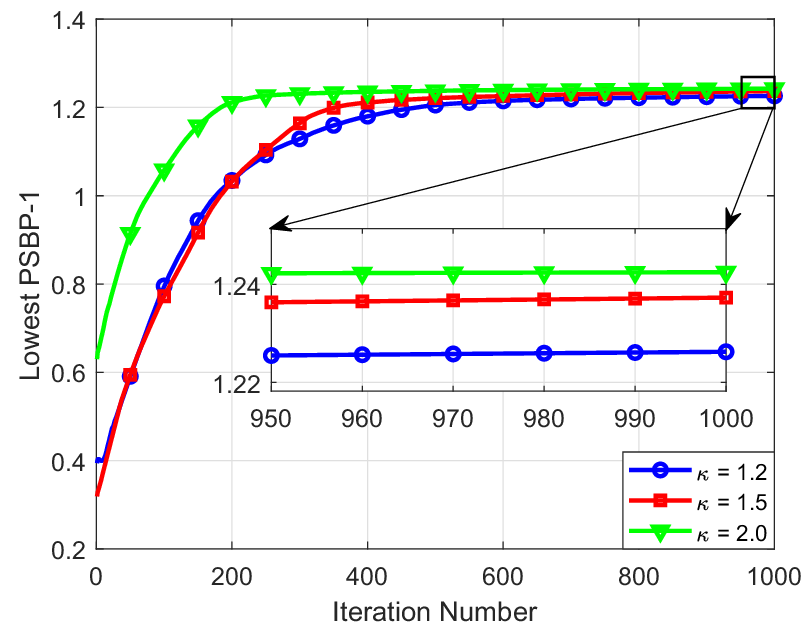} 
			\label{fig:1-1-2}
		}\hspace{-0.5em}
		\subfigure[]{
			\includegraphics[width=0.32\linewidth]{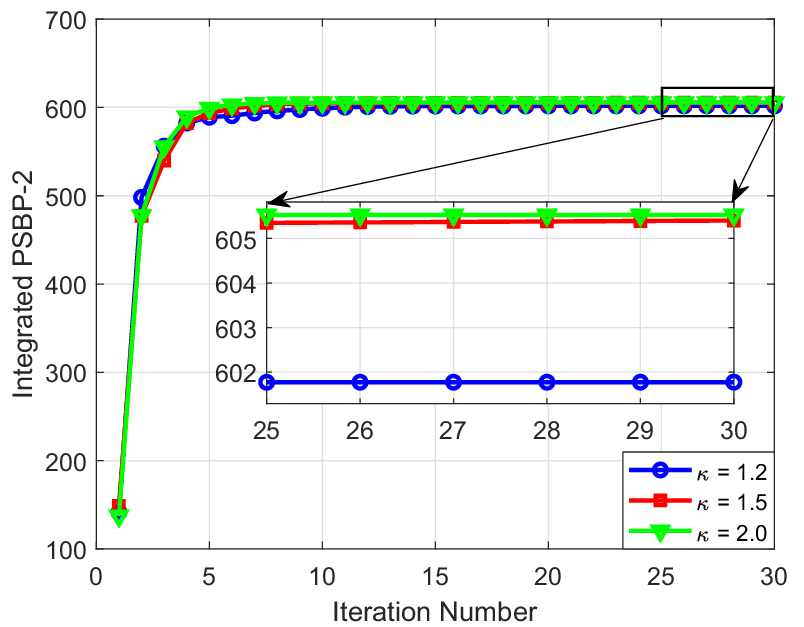}
			\label{fig:1-1-3}
		}
		\vspace{-1em}
		\caption{Convergence performance of the proposed transmit beampattern design algorithms with different PAPR requirements $\kappa = 1.2,1.5,2.0$ in Case 1-2. (a) PCRB Oriented Method; (b) PSBP-1 Oriented Method; (c) PSBP-2 Oriented Method.}
		\label{fig:1-1}
		\vspace{-1.5em}
	\end{figure*}
	
	For notation simplification, let PSBP-1 and PSBP-2 denote the specific form of PSBP \eqref{eq:22} involved in the fair PSBP design and the integrated PSBP design, respectively.
	Accordingly, their corresponding transmit beampattern design methods are denoted by the PSBP-1 and PSBP-2 oriented transmit beampattern methods, respectively. 
	
	\subsubsection{Parameter Setting}
	In the considered colocated radar system, we assume that the colocated radar is equipped with $M_t = M_r = M = 8$ transmit/receive antennas.
	The number of samples for each radar pulse is $L = 25$ and the sampling number of the angular range $\left[ { - \pi /2,\pi /2} \right]$ is $D = 361$.
	The transmit power is set as $P = 1 $, and the radar noise power is set as $\sigma _{\rm{r}}^2 = 0{\rm{dB}}$.
	The complex amplitude is assumed to be ${\left| \varsigma  \right|^2} = 0 {\rm dB}$.
	All the numerical schemes are analyzed using Matlab 2022b version and performed in a standard PC with Intel(R) CPU(TM) Core i7-10700 2.9 GHz and 16 GB RAM.
	
	\subsubsection{DoA Estimation}
	For the DoA estimation based on the transmit beampattern optimization methods exploiting target location distribution, we adopt the maximum-a-posteriori (MAP) estimation method, which is given by
	\vspace{-1em}
	\begin{equation}
	{{\hat \theta }_{{\rm{MAP}}}} = \arg \;\mathop {\max }\limits_\theta  \;\left[ {\ln \left( {f\left( {{\bf{Y}}|{\theta}} \right)} \right) + \ln \left( {f\left( {\theta} \right)} \right)} \right].
	\end{equation}
	Besides, the MSE between the estimated angle and the actual angle is defined as 
	\begin{equation}
	{\rm{MSE  =  }}\frac{1}{{\cal N}}\textstyle\sum\limits_{n \in {\cal N}} {{( {{{\hat \theta }_n} - {\theta _n}} )^2}},
	\end{equation}
	where $\mathcal{N}$ denotes Monter-Carlo trials for each angular point. ${{\theta _n}}$ and ${{{\hat \theta }_n}}$ respectively denote the actual and estimated DoA of the target.
	In the numerical simulation, we set $\mathcal{N} = 10^4$.
	
	\subsubsection{Benchmarks}
	To demonstrate the superiority of the proposed algorithms and verify the effectiveness of target location distribution exploitation methods, we include the following benchmarks.
	\begin{enumerate}
		\item  {\textbf{SCA\&IP Based Algorithms.}
			To demonstrate the superiority of the ADMM-based solutions to $\mathcal{P}_{\mathrm{PCRB},\mathbf{X}}^{1-1}$, $\mathcal{P}_{\mathrm{PSBP},\mathbf{X}}^{2-1}$, $\mathcal{P}_{\mathrm{PSBP},\mathbf{X}}^{3-1}$, we compare the proposed methods with SCA\&IP-based algorithms using successive convex approximation (SCA) to transform the original non-convex problem to a convex one and interior-point method to tackle the reformulated problem.}
		\item \textbf{CRB Oriented transmit beampattern.} 
		In this benchmark, the radar transmit beampattern is optimized for a deterministic angle $\theta_0$. 
		With other system parameters kept the same, we assume the angular location of the target is $\theta_0 \in {\bf \Theta}$.
		\item \textbf{Omnidirectional Radar transmit beampattern.}
		The radiation or reception pattern that is uniform or nearly uniform in all directions is a widely used typical radar waveform.
	\end{enumerate}

	\vspace{-1em}
	\subsection{Scenario 1}\label{sec:7-1}

	In this scenario, for the target location distribution, we assume there is $K = 1$ possible angular location interval with the probability $p = 1$.
	Based on the width of the target occurrence regions, we consider four cases:\\
	\textbf{Case 1-1:} $ {\bf \Theta} \!\!=\!\! \left[ -\pi/36, \pi/36 \right]$; \textbf{Case 1-2:} $ {\bf \Theta}\!\! = \!\!\left[ -\pi/18, \pi/18 \right]$;
	\textbf{Case 1-3:} $ {\bf \Theta} = \left[ -\pi/9, \pi/9 \right]$; \textbf{Case 1-4:} $ {\bf \Theta} = \left[ -\pi/6, \pi/6 \right]$.
	
	\subsubsection{Convergence Performance of Algorithm}
	
	In Fig. \ref{fig:1-1}, we present the convergence performance of the proposed transmit beampattern design algorithms via PCRB, PSBP-1, and PSBP-2 oriented methods, with different PAPR requirements $\kappa = 1.2, 1.5, 2.0$ for \textbf{Case 1-2}.
	Fig. \ref{fig:1-1-1}, Fig. \ref{fig:1-1-2} and Fig. \ref{fig:1-1-3} plot the PCRB, the lowest PSBP-1 and the integrated PSBP-2 versus the iteration number with different $\kappa$, respectively.
	It can be observed that with the iteration number increasing, the values of PCRB, the lowest PSBP-1, and integrated PSBP-2 reach stationary, validating the convergency of the proposed algorithms.
	Moreover, with higher PAPR $\kappa$, the PCRB becomes lower, while the lowest PSBP-1 and integrated PSBP-2 become higher.
	This validates that the higher DoF introduced by the higher PAPR can improve radar performance.

	\subsubsection{Amplitude of Designed Waveform with Different PAPR}
	
	\begin{figure}[!t]
		\centering  
		\subfigure[]{
			\label{fig:1-2-1}
			\includegraphics[width=0.7\linewidth]{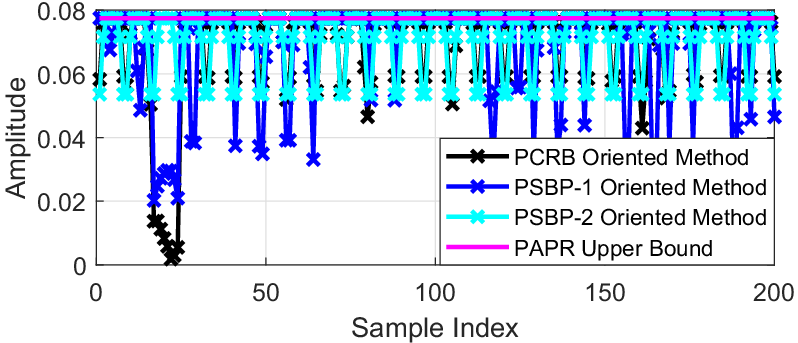}}
		
		\vspace{-1em}
		\subfigure[]{
			\label{fig:1-2-2}
			\includegraphics[width=0.7\linewidth]{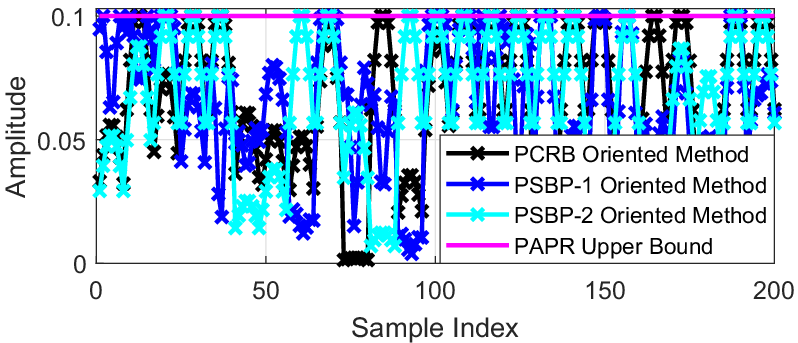}}
		\vspace{-0.5em}	
		\caption{The amplitude of the designed waveform proposed methods with different PAPR requirements in Case 1-2. (a) PAPR requirement $\kappa = 1.2$. (b) PAPR requirement $\kappa = 2.0$.}
		\label{fig:1-2}
		\vspace{-1.8em}
	\end{figure}
	
	In Fig. \ref{fig:1-2}, we show the amplitude of the radar waveform designed by the proposed three methods.
	The sample index is defined as $l\times m,\forall l = 1,\dots,L,m = 1,\dots,M_t$.
	Specifically, Figs. \ref{fig:1-2-1}, \ref{fig:1-2-2} plot the amplitude of the waveform with the PAPR requirement $\kappa = 1.2$ and $\kappa = 2.0$, respectively.
	It can be seen that the waveform amplitudes obtained by all the proposed methods are not higher than the PAPR upper bound.
	These results demonstrate the effectiveness of the proposed methods in ensuring that the waveform PAPR complies with hardware specifications in the practical scenario.
	
	\subsubsection{Beampattern Performance}

\begin{figure}[t]
		\centering
		\subfigure[]{
			\includegraphics[width=0.48\linewidth]{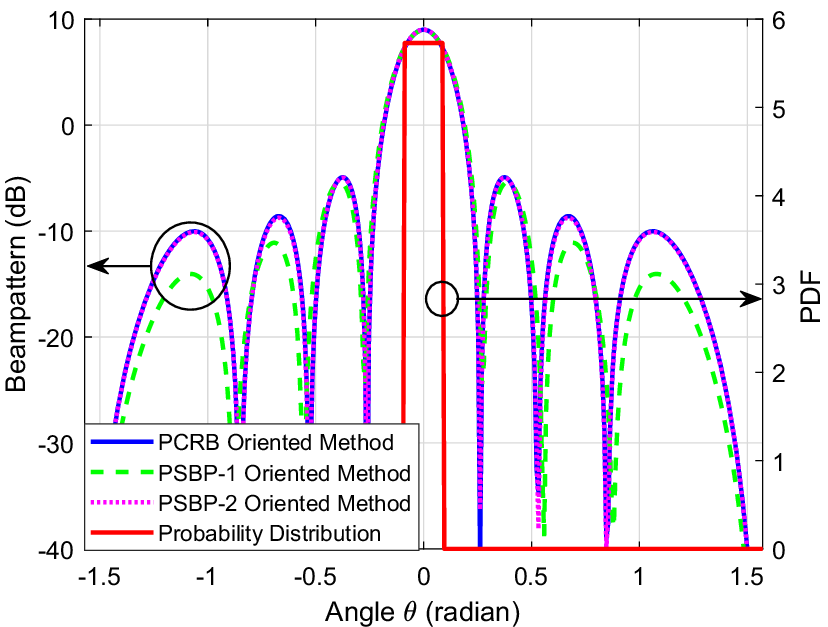} 
			\label{fig:1-3-1}
		}\hspace{-1em}
		\subfigure[]{
			\includegraphics[width=0.48\linewidth]{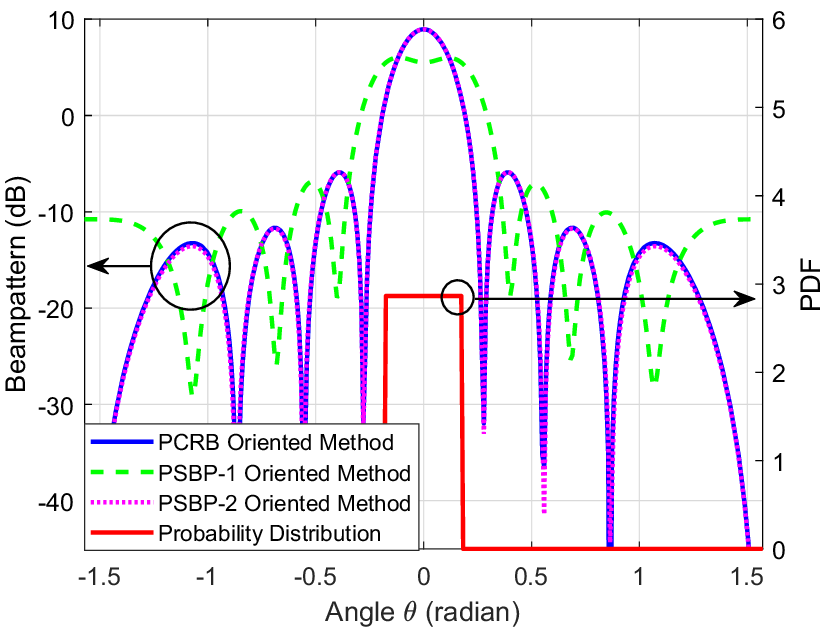} 
			\label{fig:1-3-2}
		}\vspace{-1em}
		
		\subfigure[]{
			\includegraphics[width=0.48\linewidth]{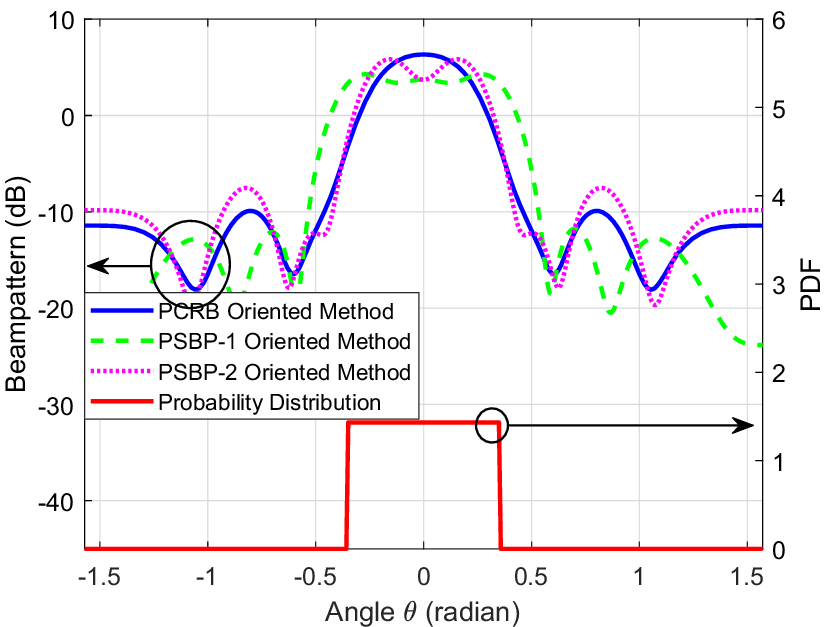}
			\label{fig:1-3-3}
		}\hspace{-1em}
		\subfigure[]{
			\includegraphics[width=0.48\linewidth]{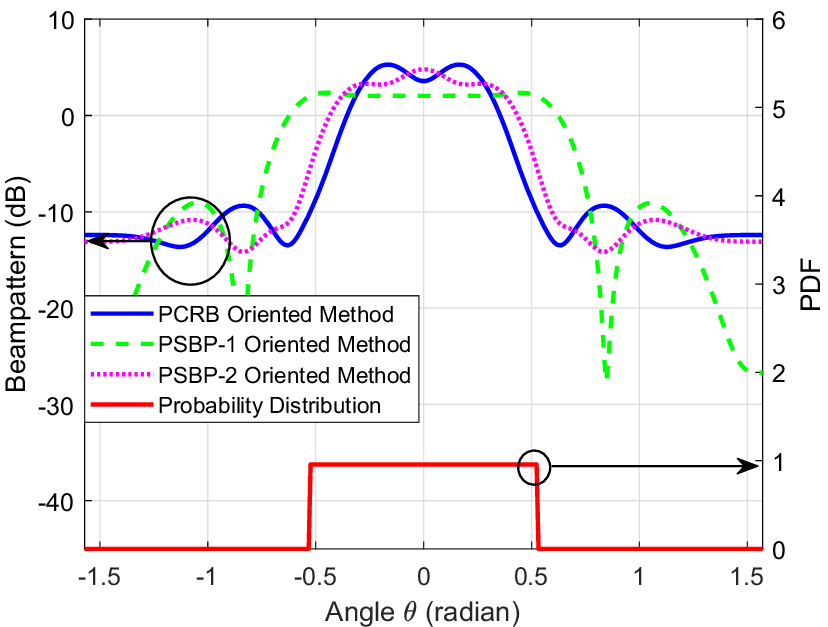}
			\label{fig:1-3-4}
		}
		\vspace{-1em}
		\caption{Comparisons of the radar beampattern obtained by the proposed methods in different target distribution cases. (a) Case 1-1; (b) Case 1-2; (c) Case 1-3; (d) Case 1-4.}
		\label{fig:1-3}
		\vspace{-1.5em}
	\end{figure}

	In Fig. \ref{fig:1-3}, we compare the radar beampattern obtained by the proposed three methods in different distribution cases.
	When the width of the target distribution region is small as shown in Fig. \ref{fig:1-3-1}, the beampatterns generated by all methods are almost the same, with high mainlobe levels in the high-probability target occurrence region.
	As the location uncertainty width gets larger from Fig. \ref{fig:1-3-1} to Fig. \ref{fig:1-3-4}, we observe that the PSBP-1 oriented method can better maintain a satisfactory high mainlobe level over the whole high-probability target occurrence region, while the PCRB and PSBP-2 oriented methods can guarantee a relatively higher level on the narrow central region at a cost of level reduction on the edge of the target distribution region.
	This shows that the PSBP-1 oriented method realizes a better beampattern performance than others when the target location uncertainty is large.
	
	\subsubsection{DoA Estimation Performance}

\begin{figure}[t]
		\centering
		\subfigure[]{
			\includegraphics[width=0.46\linewidth]{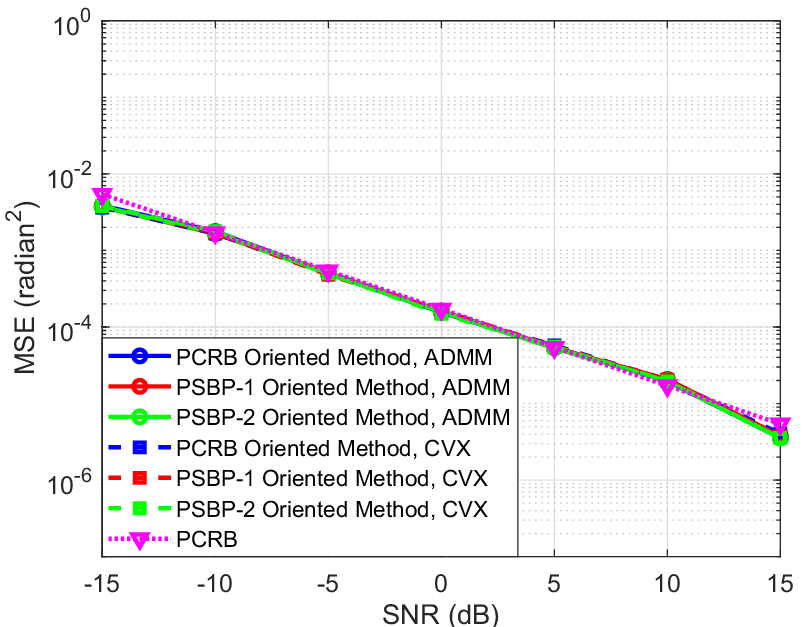} 
			\label{fig:1-4-1}
		}\hspace{-1em}
		\subfigure[]{
			\includegraphics[width=0.46\linewidth]{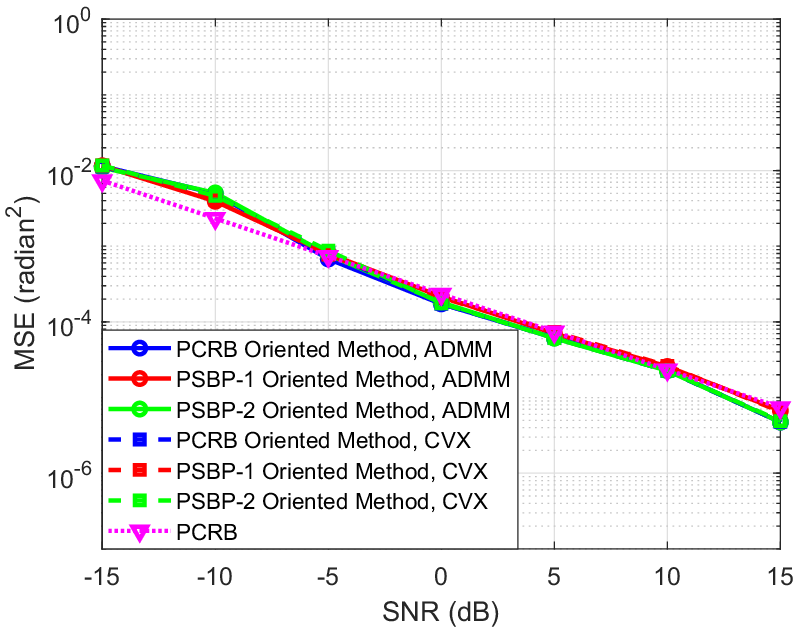} 
			\label{fig:1-4-2}
		}\vspace{-1em}
		
		\subfigure[]{
			\includegraphics[width=0.46\linewidth]{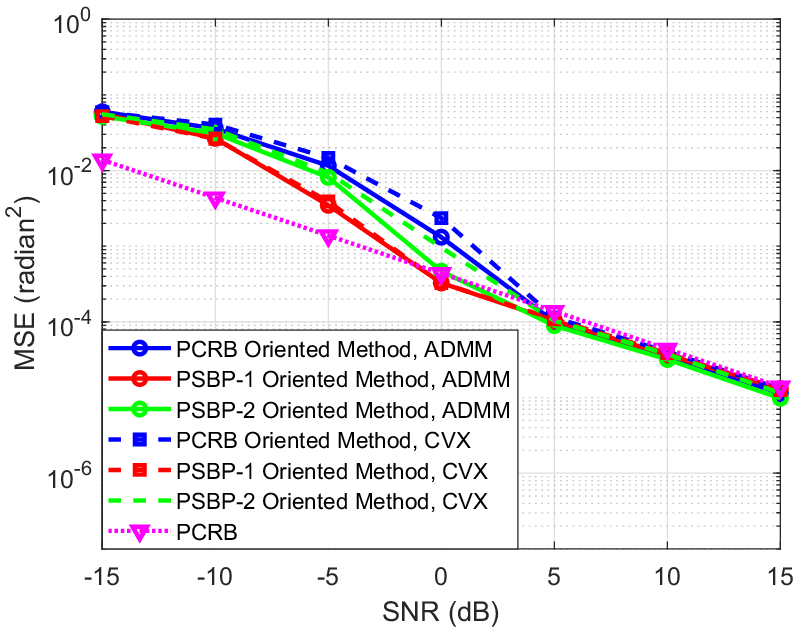}
			\label{fig:1-4-3}
		}\hspace{-1em}
		\subfigure[]{
			\includegraphics[width=0.46\linewidth]{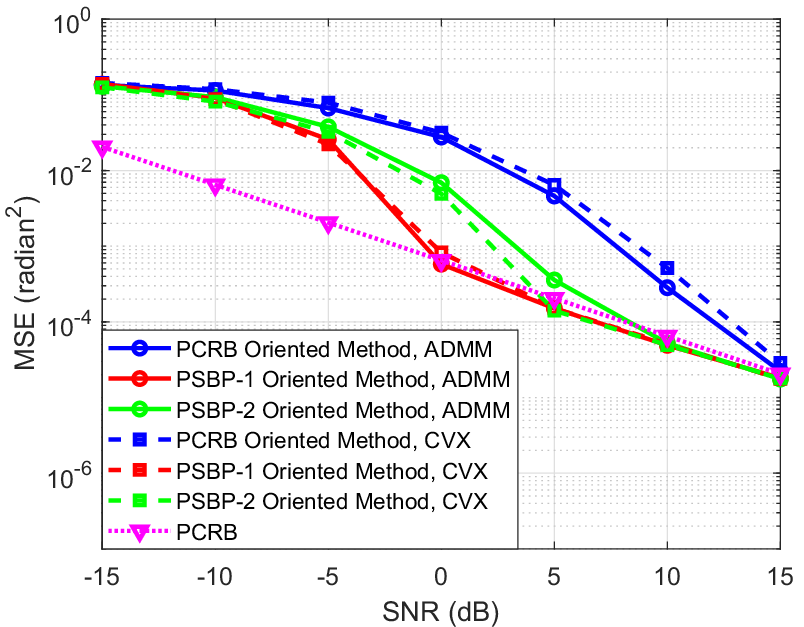}
			\label{fig:1-4-4}
		}
		\vspace{-0.5em}
		\caption{Comparisons of DoA estimation performance by the proposed ADMM-based algorithms and the SCA\&IP-based algorithms in different target distribution cases: the MSE versus SNR. (a) Case 1-1; (b) Case 1-2; (c) Case 1-3; (d) Case 1-4.}
		\label{fig:1-4}
		\vspace{-1.5em}
	\end{figure}

	In Fig. \ref{fig:1-4}, we compare the DoA estimation performance of the transmit beampattern designed by the proposed ADMM-based algorithms and  {SCA\&IP}-based algorithms in different target distribution cases by plotting the MSE versus radar SNR.
	It can be seen in Fig. \ref{fig:1-4-1} that the curves of MSEs obtained by all the algorithms are nearly superimposable on the curve of the theoretical PCRB in Case 1-1.
	As the width of target distribution regions gets larger, the MSE curves get away from the PCRB curves, as shown from Fig. \ref{fig:1-4-2} to Fig. \ref{fig:1-4-4}.
	Specifically, in the case of wider target distribution regions, the MSE curves of PSBP-1 oriented method can reach the PCRB bound at a relatively lower SNR, followed by the PSBP-2 oriented method, while the MSE curves of PCRB oriented method approach the PCRB curves at a higher SNR.
	These results show that in a single-occurrence-region scenario, the DoA estimation performance of the transmit beampattern designed by the PSBP-1 oriented algorithm is the best among the proposed three algorithms, followed by the PSBP-2 oriented algorithm, and the PCRB oriented algorithm in the last position.
	We can also notice that MSE curves of the proposed ADMM-based algorithms are close to those of the  {SCA\&IP}-based algorithms.
	This validates the effectiveness of the proposed ADMM-based transmit beampattern design algorithms.

	\subsubsection{Computation Time}
	
	Table \ref{tab:1} presents the computation time of the proposed ADMM-based algorithms and  {SCA\&IP}-based algorithms for three transmit beampattern design methods in different cases.
	Generally, it can be seen that the computation time of the proposed ADMM-based algorithms is considerably lower than the  {SCA\&IP}-based algorithms.
	This indicates that the proposed ADMM-based algorithms have great potential to realize real-time computation in practice.
	Moreover, for the ADMM-based algorithm, we notice that the PSBP-2 oriented algorithm takes nearly 0.1 seconds, the PCRB oriented algorithm takes 1-2 seconds, and the PSBP-1 oriented algorithm takes 10-30 seconds.
	
	\begin{table}[!h]
		\vspace{-1em}
		\centering
		\scriptsize
		\begin{center}
			\caption{Computation time of the proposed ADMM-based algorithms and  {SCA\&IP}-based algorithms (unit: second)}
			\label{Table1}
			\vspace{-0.5em}
			\begin{tabular}{|c|c|c|c|c|c|}
				\hline
				Method                                                                     &      & Case 1-1 & Case 1-2 & Case 1-3 & Case 1-4 \\ \hline
				\multirow{2}{*}{\begin{tabular}[c]{@{}c@{}}PCRB\\ Oriented\end{tabular}}   & ADMM & 1.812    & 1.728     & 2.103     & 1.757     \\ \cline{2-6} 
				&  {SCA\&IP}  & 940.017   & 1170.235  & 1459.215  & 1694.736  \\ \hline
				\multirow{2}{*}{\begin{tabular}[c]{@{}c@{}}PSBP-1\\ Oriented\end{tabular}} & ADMM & 10.273    & 14.144    & 21.754   & 29.346    \\ \cline{2-6} 
				&  {SCA\&IP}  & 797.727   & 845.498   & 878.421   & 905.318   \\ \hline
				\multirow{2}{*}{\begin{tabular}[c]{@{}c@{}}PSBP-2\\ Oriented\end{tabular}} & ADMM & 0.120     & 0.123    & 0.145     & 0.152     \\ \cline{2-6} 
				&  {SCA\&IP}  & 844.462   & 881.488   & 882.686   & 884.310   \\ \hline
			\end{tabular}\label{tab:1}
		\end{center}
		\vspace{-2em}
	\end{table}
	
	Combining Figs. \ref{fig:1-1}-\ref{fig:1-4}  and Table \ref{tab:1}, we provide insights that, although PSBP-1 can achieve better beampattern and DoA estimation, the design time cost of PSBP-1 is relatively higher than the other two methods. 
	This suggests that every target location distribution exploitation method has its pros and cons. 
	Therefore, it is important to choose a more suitable transmit beampattern design algorithm when considering the trade-off between computation time requirements and radar performance in practice.
	
	\vspace{-1em}
	\subsection{Scenario 2}
	
	\begin{figure}[!t]
		\centering
		\subfigure[]{
			\includegraphics[width=0.48\linewidth]{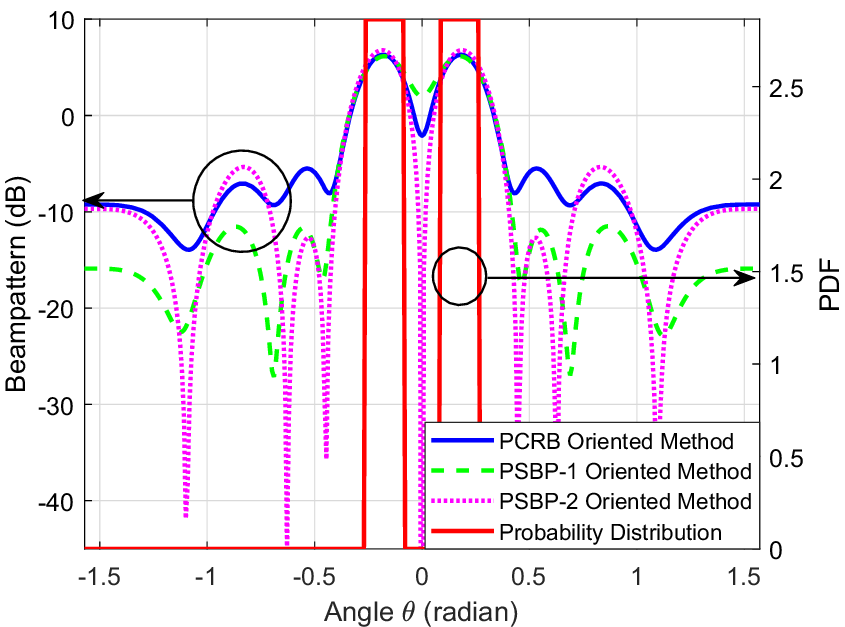} 
			\label{fig:2-1-1}
		}\hspace{-1em}
		\subfigure[]{
			\includegraphics[width=0.48\linewidth]{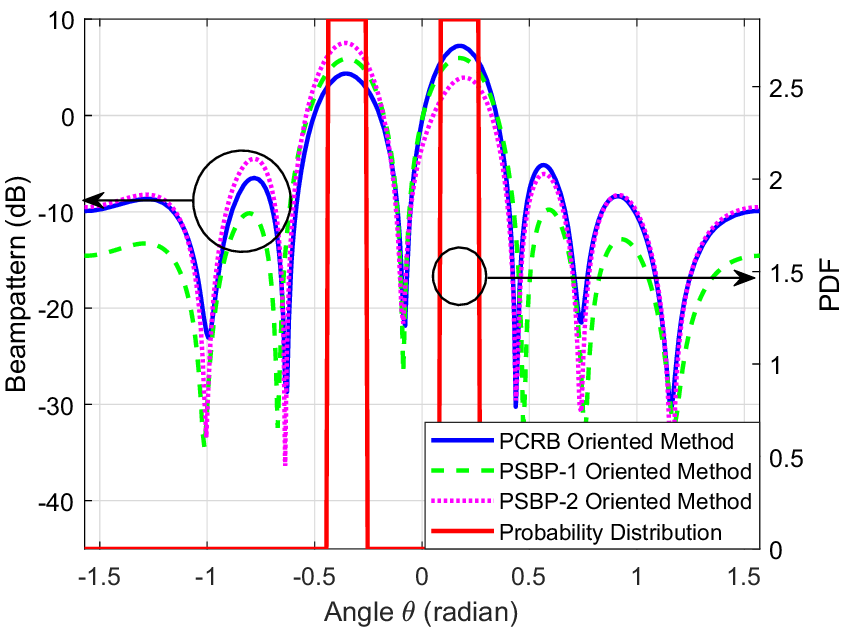} 
			\label{fig:2-1-2}
		}\vspace{-1em}
		
		\subfigure[]{
			\includegraphics[width=0.48\linewidth]{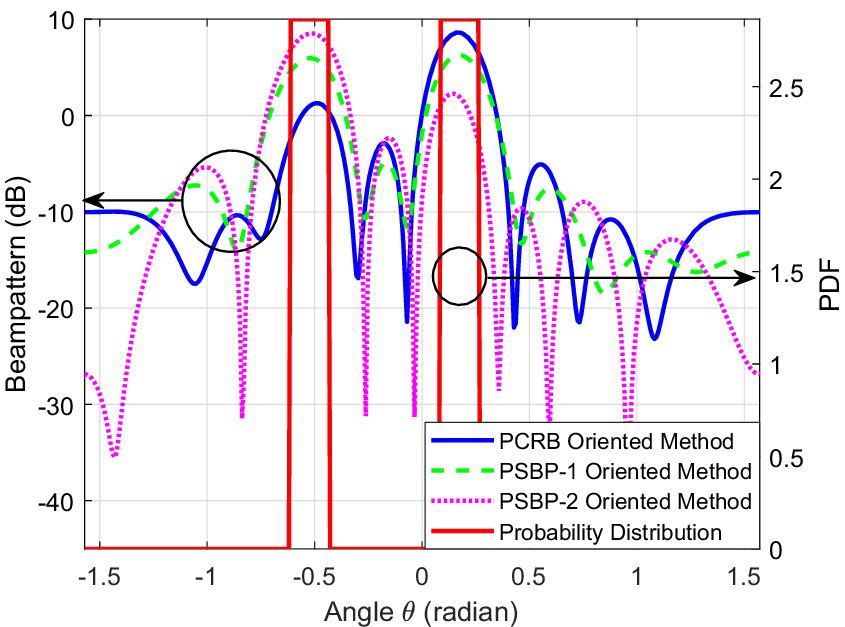}
			\label{fig:2-1-3}
		}\hspace{-1em}
		\subfigure[]{
			\includegraphics[width=0.48\linewidth]{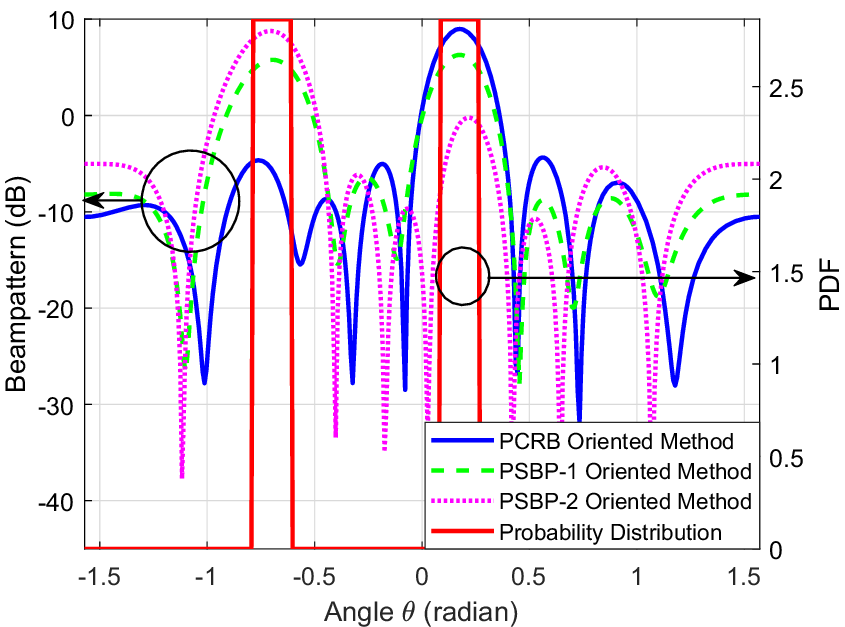}
			\label{fig:2-1-4}
		}\vspace{-1em}
		
		\subfigure[]{
			\includegraphics[width=0.48\linewidth]{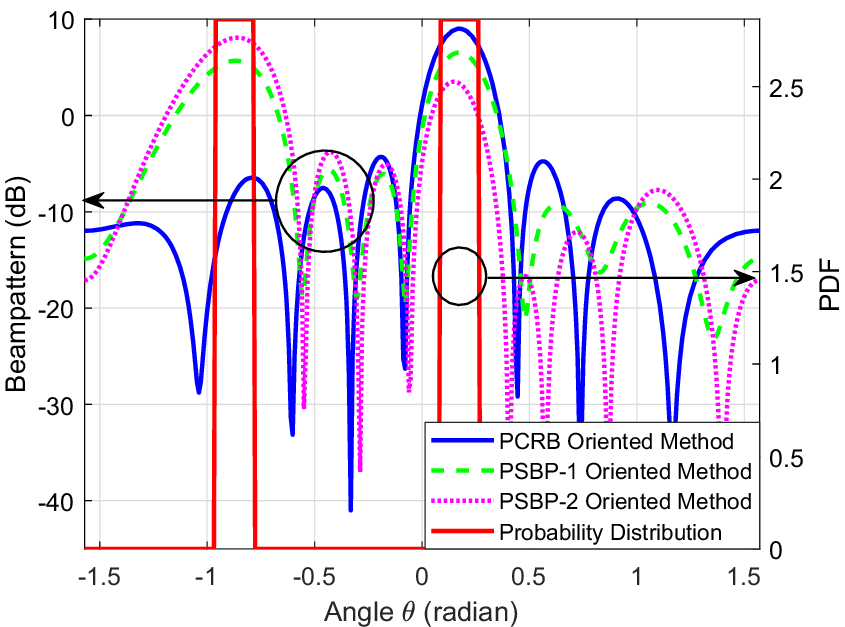}
			\label{fig:2-1-5}
		}\hspace{-1em}
		\subfigure[]{
			\includegraphics[width=0.48\linewidth]{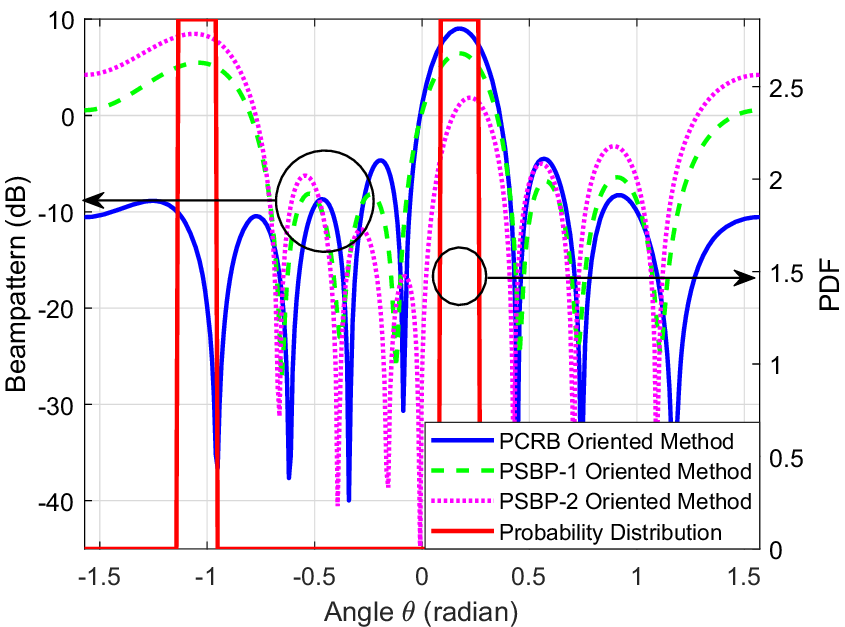}
			\label{fig:2-1-6}
		}
		\vspace{-1em}
		\caption{Comparisons of the radar beampattern obtained by the proposed methods in different target distribution cases. (a) Case 2-1; (b) Case 2-2; (c) Case 2-3; (d) Case 2-4; (e) Case 2-5; (f) Case 2-6.}
		\label{fig:2-1}
		\vspace{-1.5em}
	\end{figure}
	
	In this scenario, for target location distribution, we assume there are $K = 2$ possible angular location intervals with the probability $p_1 = p_2 = 0.5$.
	Specifically, based on the separation width of target distribution intervals, we consider six cases:\\
	\textbf{Case 2-1:} ${\bf{\Theta }} = \left[ { - 3\pi /36,- \pi /36} \right] \cup \left[ {\pi /36, 3\pi /36} \right]$;\\
	\textbf{Case 2-2:} ${\bf{\Theta }} = \left[ { -  5\pi/36, - 3 \pi /36} \right] \cup \left[ {\pi /36, 3\pi /36} \right]$;\\
	\textbf{Case 2-3:} ${\bf{\Theta }} = \left[ { - 7\pi/36, - 5\pi/36} \right] \cup \left[ {\pi /36, 3\pi /36} \right]$;\\
	\textbf{Case 2-4:} ${\bf{\Theta }} = \left[ { - 9\pi /36, - 7\pi /36} \right] \cup \left[ {\pi /36,3\pi /36} \right]$;\\
	\textbf{Case 2-5:} ${\bf{\Theta }} = \left[ { - 11\pi /36, - 9\pi /36} \right] \cup \left[ {\pi /36,3\pi /36} \right]$;\\
	\textbf{Case 2-6:} ${\bf{\Theta }} = \left[ { - 13\pi /36, - 11\pi /36} \right] \cup \left[ {\pi /36,3\pi /36} \right]$.
	
	\subsubsection{Beampattern Performance}
	
	Fig. \ref{fig:2-1} presents the radar beampatterns obtained by the proposed methods from Case 2-1 to Case 2-6.
	It can be seen that both methods can achieve relatively higher mainlobe levels in the angular regions with high target occurrence probability.
	Specifically, the PCRB oriented method can realize a higher mainlobe level on the angular interval that is closer to the central angle, albeit with a sacrifice in the mainlobe levels on the other angular interval, while the PSBP-2 oriented method has the opposite effect.
	We can also observe that the mainlobe levels obtained by the PSBP-1 oriented method are guaranteed to be high on both target distribution intervals.
	This phenomenon becomes more obvious with the separation width increasing, as presented from Fig. \ref{fig:2-1-1} to Fig. \ref{fig:2-1-6}.
	The results show that the PCRB oriented method can improve the mainlobe level on the angular intervals closer to the central angle, the PSBP-2 oriented method can improve the mainlobe level on the angular intervals far away from the central angle, and the PSBP-2 oriented method has a better mainlobe level control on each target distribution interval.

	\begin{figure}[!t]
		\centering  
		\includegraphics[width=0.7\linewidth]{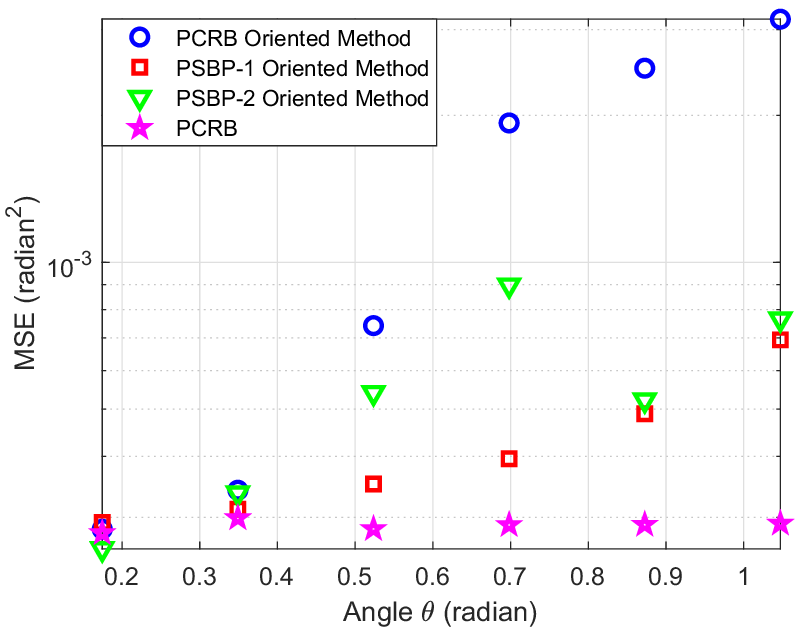}
		\vspace{-0.5em}
		\caption{Comparisons of radar DoA estimation performance in different target distribution cases (From Case 2-1 to Case 2-6): the MSE of the DoA estimation versus the separation width.}
		\label{fig:2-2}
		\vspace{-1em}
	\end{figure}
	
	\subsubsection{DoA Estimation Performance}
	
	In Fig. \ref{fig:2-2}, we compare the radar DoA estimation performance of the proposed methods in different target distribution cases (from Case 2-1 to Case 2-6) by plotting MSEs versus the separation width.
	It can be seen that MSEs of DoA estimation increase with the separation width increasing and MSEs become more distant from the PCRB with increasing separation width, except for the PSBP-2 oriented method at the last two points.
	We can also observe that the MSE of DoA estimation obtained by the PSBP-1 oriented method is closer to the PCRB, followed by the PSBP-2 oriented method, and the PCRB oriented method in the last position.
	Generally speaking, in the multiple-occurrence-region scenario, the results indicate that the DoA estimation performance of the proposed methods degrades when the separation width becomes larger, and the DoA performance of the PSBP-1 oriented method is the best and the PCRB oriented method is the worst.
	
	\subsubsection{Impact of Number of Antennas on Beampattern Performance}
	
	\begin{figure}[!t]
		\centering  
		\includegraphics[width=0.75\linewidth]{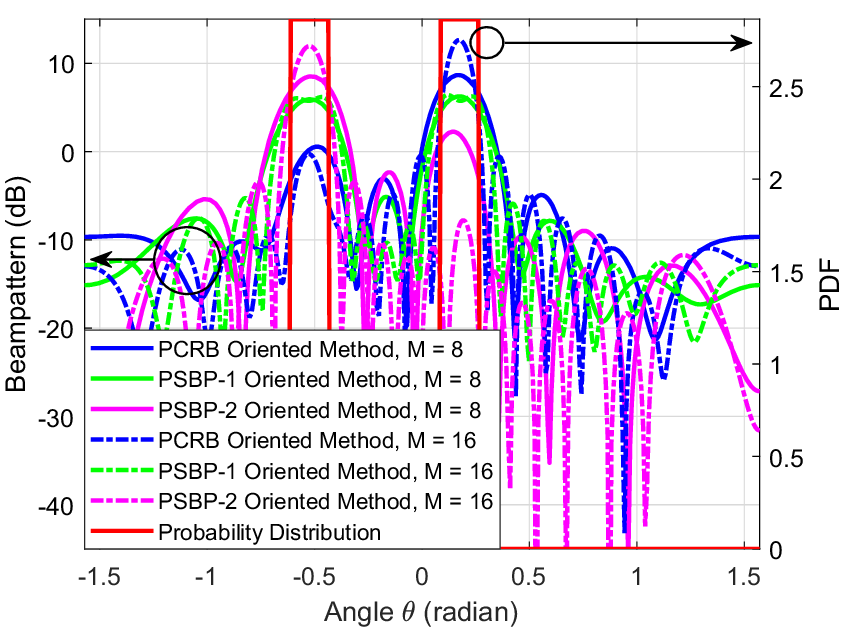}
		\vspace{-0.5em}
		\caption{Beampattern obtained by the proposed methods with a different number of transmit/receive antennas $M = 8,16$ in Case 2-3.}
		\label{fig:2-3}
		\vspace{-1.5em}
	\end{figure}
	
	In Fig. \ref{fig:2-3}, we present the radar transmit beampattern obtained by the proposed methods, with a different number of transmit/receive antennas $M=8,16$, in \textbf{Case 2-3}.
	We can observe that with the number of antennas increasing, the PCRB oriented method improves mainlobe level on the angular intervals closer to the central angle, while the PSBP-2 oriented method improves mainlobe level on the angular intervals far away from the central angle, and the PSBP-1 oriented method can better maintain a high mainlobe level among the whole region with high probability.
	These results indicate that it is necessary to choose a suitable number of transmit/receive antennas to enhance the radar beampattern behavior.

	\begin{figure}[!t]
		\centering  
		\includegraphics[width=0.7\linewidth]{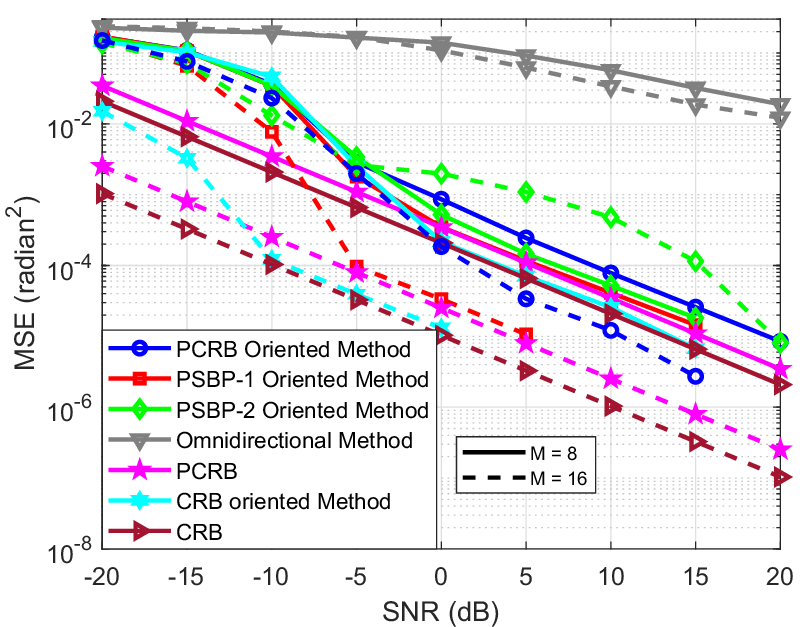}
		\vspace{-0.5em}
		\caption{Comparisons of radar DoA estimation performance with a different number of transmit/receive antennas $M = 8, 16$: the MSE of the DoA estimation versus the radar SNR.}
		\label{fig:2-4}
		\vspace{-0.5em}
	\end{figure}
	
	\begin{figure}[!t]
		\centering
		\subfigure[]{
			\includegraphics[width=0.48\linewidth]{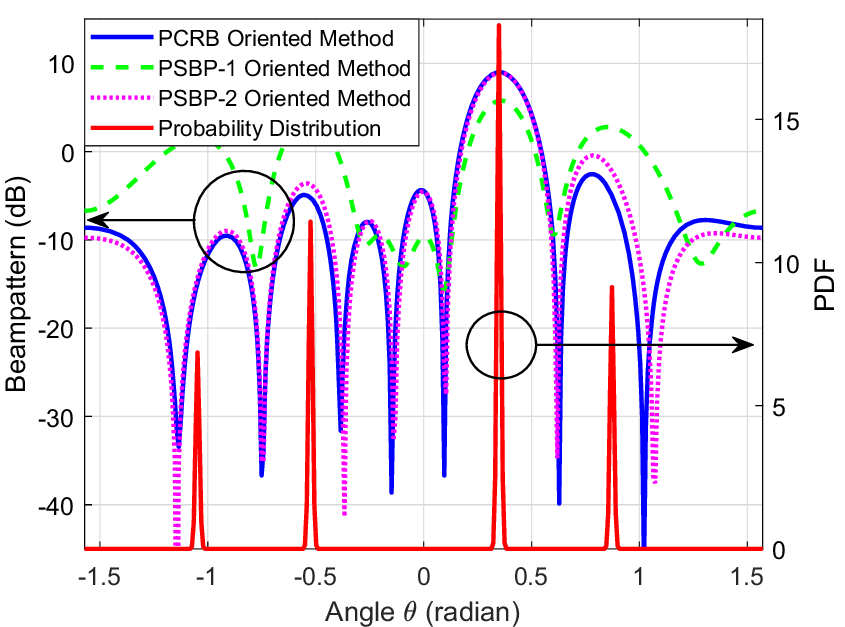} 
			\label{fig:3-1-1}
		}\hspace{-1em}
		\subfigure[]{
			\includegraphics[width=0.48\linewidth]{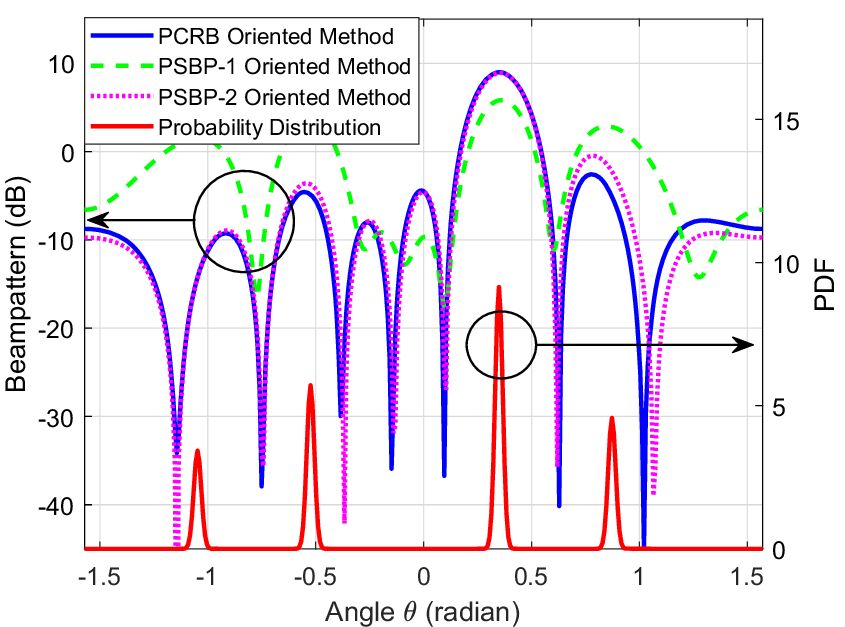} 
			\label{fig:3-1-2}
		}\vspace{-1em}
		
		\subfigure[]{
			\includegraphics[width=0.48\linewidth]{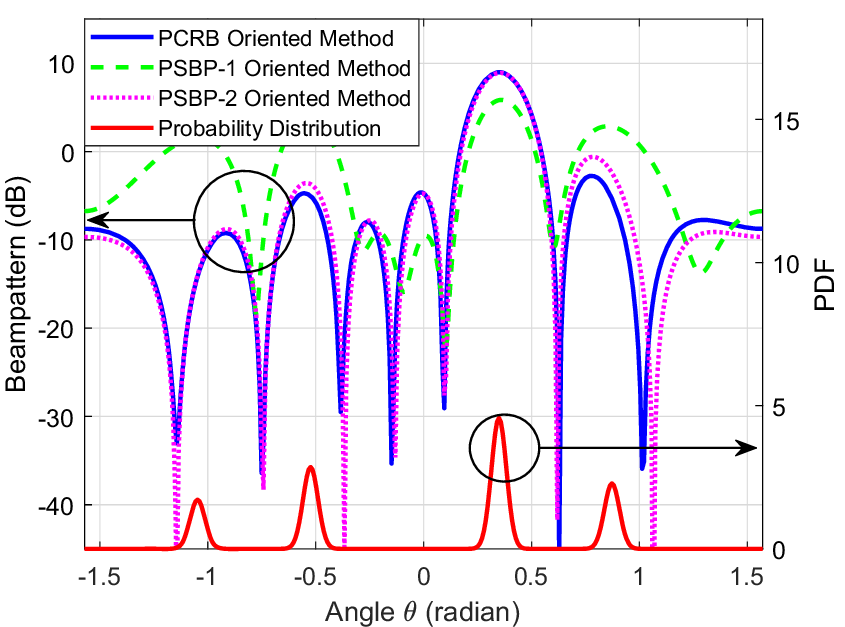}
			\label{fig:3-1-3}
		}\hspace{-1em}
		\subfigure[]{
			\includegraphics[width=0.48\linewidth]{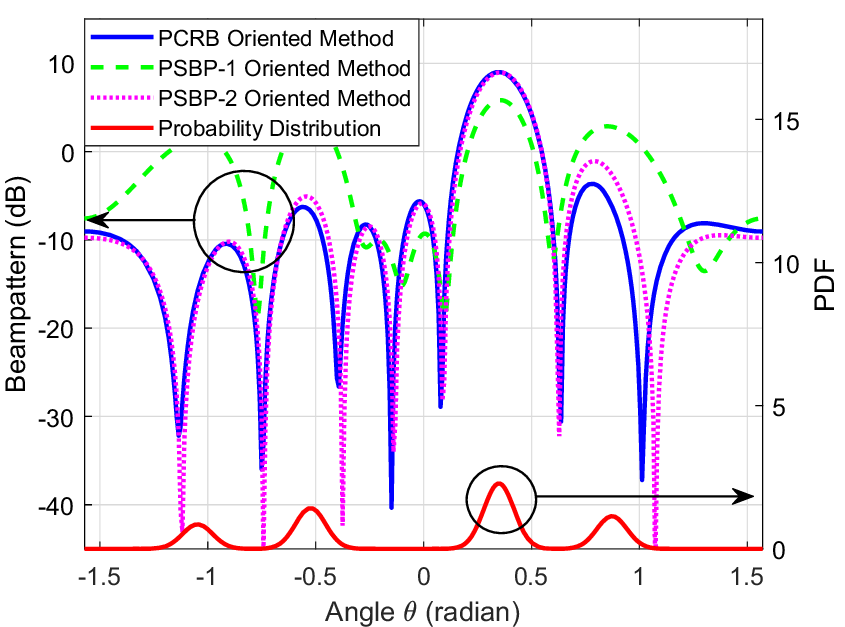}
			\label{fig:3-1-4}
		}
		\vspace{-1em}
		\caption{Comparisons of the radar beampattern obtained by the proposed methods with different standard deviation $\sigma _\theta$. (a) ${\sigma _\theta = \pi/360}$ ; (b) ${\sigma _\theta = \pi/180}$; (c) ${\sigma _\theta = \pi/90}$; (d) ${\sigma _\theta = \pi/45}$.}
		\label{fig:3-1}
		\vspace{-1.5em}
	\end{figure}
	
	\subsubsection{Impact of Number of Antennas to DoA estimation Performance}
	In Fig. \ref{fig:2-4}, we present the MSE of the DoA estimation versus the radar SNR for different transmit beampattern optimization methods with the number of antennas $M = 8,16$.
	It can be seen that the MSEs of DoA estimation obtained by all the methods are decreasing with the radar SNR increasing, and are bounded by the corresponding PCRB/CRB.
	With the number of antennas increasing, the PCBR/CRB and the MSE obtained by the PSBP-1 oriented, PCRB oriented and CRB oriented methods become lower, while the MSEs obtained by the PSBP-2 oriented methods become higher.
	Specifically, with $M = 16$, the MSEs obtained by the PSBP-1 oriented method are the closest to the PCRB, followed by the PCRB oriented method, and the PSBP-2 oriented method are the farthest from the PCRB.
	Besides, the MSEs of DoA estimation obtained by the CRB oriented are always close to the CRB with a high enough SNR.
	This indicates that a higher number of antennas improves the DoA estimation performance of the PSBP-1, PCRB, and CRB oriented methods but degrades the the DoA estimation performance of the PSBP-2 oriented methods.
	Additionally, the proposed methods can realize significantly lower MSE than the omnidirectional method.
	This demonstrates that the proposed transmit beampattern optimization methods can achieve satisfactory DoA estimation performance by exploiting the target location distribution information.
	
	\vspace{-1em}
	\subsection{Scenario 3}
	
	In this scenario, for target location distribution, we assume there are $K = 4$ possible angular points, i.e., $\theta_1 = -\pi/3,~\theta_2 = -\pi/6,~\theta_3 = \pi/9,~\theta_4 = 5\pi/18$, with probabilities $p_1 = 0.15,~p_2 = 0.25,~p_3 = 0.4,~p_4 = 0.2$.

	\subsubsection{Beampattern Performance}
	
	Fig. \ref{fig:3-1} compares the radar beampattern obtained by the proposed methods with different standard deviation, $\sigma _\theta = \pi/360, \pi/180, \pi/90, \pi/45$.
	It can be seen from Fig. \ref{fig:3-1-1} to Fig. \ref{fig:3-1-4} that with $\sigma _\theta$ increasing, the beampatterns obtained by the proposed methods do not have an obvious change. 
	Specifically, the PCRB oriented and PSBP-2 oriented methods generate almost the same beampattern, whose radiation power is not well proportional to the target occurrence probability at certain angular locations.
	The PSBP-1 oriented generates beampattern with better probability-proportional mainlobe levels on angular points $\theta_1 = -\pi/3, \theta_2 = -\pi/6, \theta_3 = \pi/9, \theta_4 = 5\pi/18$.
	This shows that the PSBP-1 oriented method control levels for the radar beampattern better than the other two methods.
	
	\subsubsection{DoA Estimation Performance}
	
	\begin{figure}[t]
		\centering
		\subfigure[]{
			\includegraphics[width=0.46\linewidth]{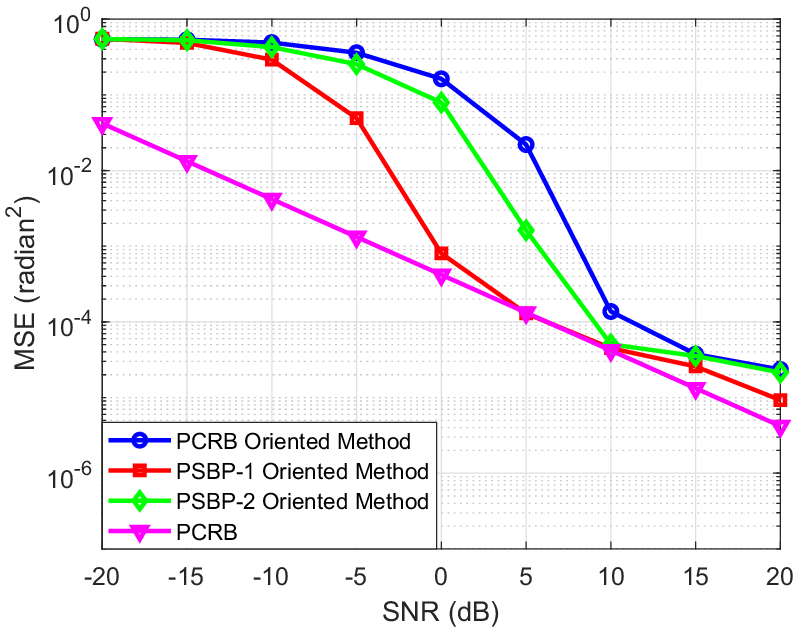} 
			\label{fig:3-2-1}
		}\hspace{-1em}
		\subfigure[]{
			\includegraphics[width=0.46\linewidth]{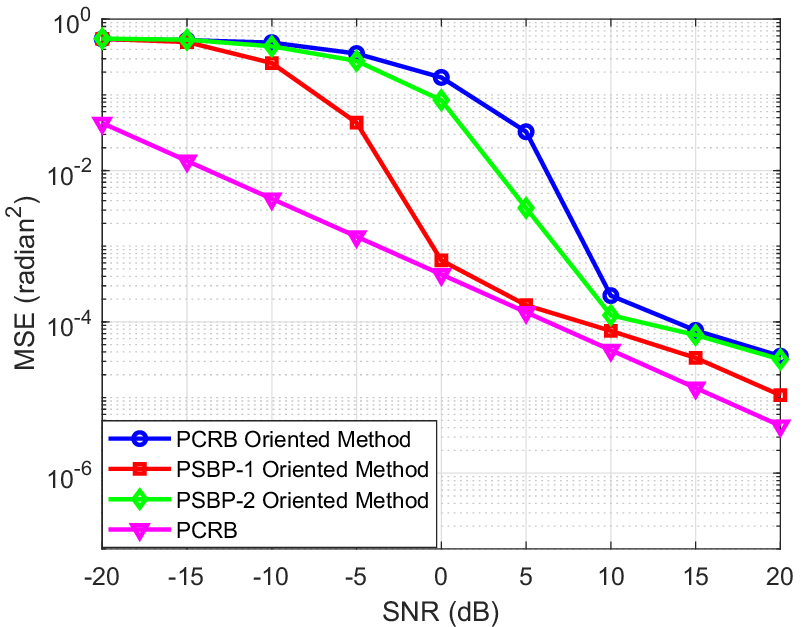} 
			\label{fig:3-2-2}
		}\vspace{-1em}
		
		\subfigure[]{
			\includegraphics[width=0.46\linewidth]{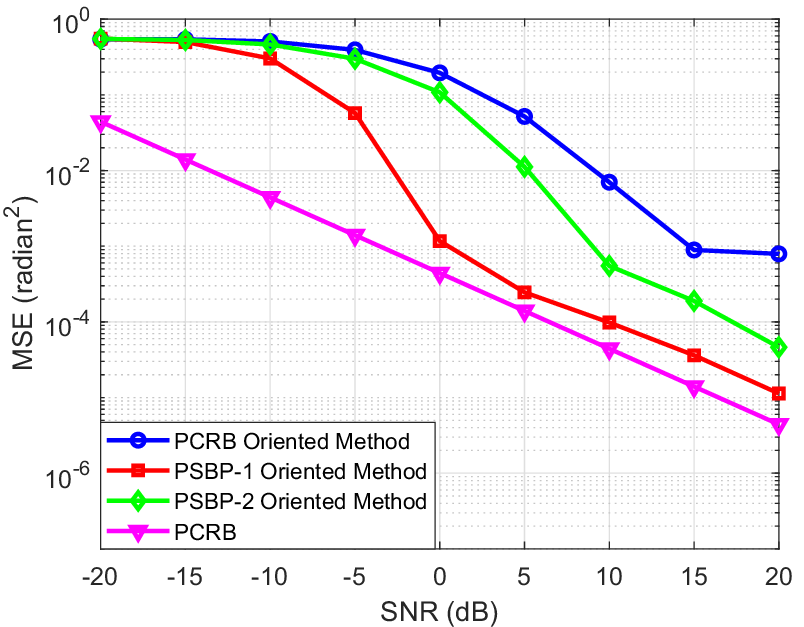}
			\label{fig:3-2-3}
		}\hspace{-1em}
		\subfigure[]{
			\includegraphics[width=0.46\linewidth]{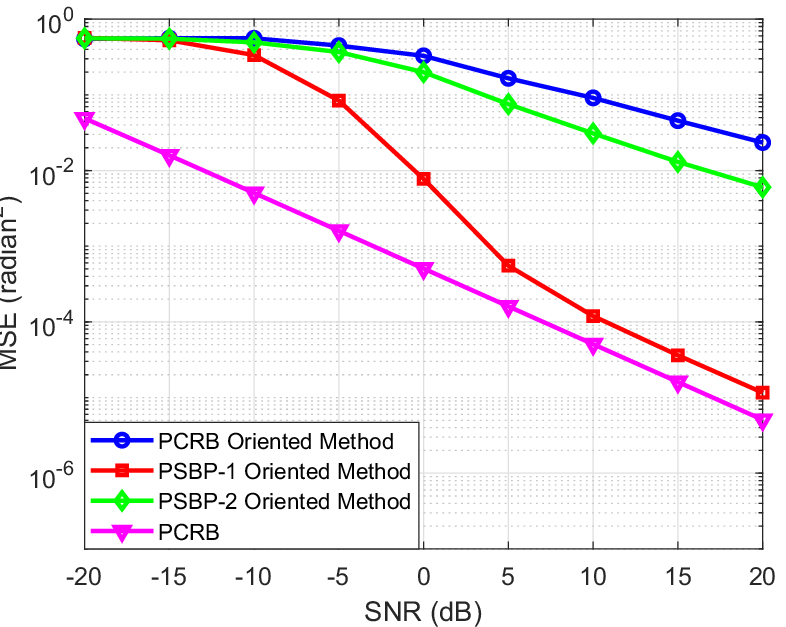}
			\label{fig:3-2-4}
		}
		\vspace{-1em}
		\caption{Comparisons of the DoA estimation performance by the proposed methods with different standard deviation $\sigma _\theta$. (a) ${\sigma _\theta = \pi/360}$ ; (b) ${\sigma _\theta = \pi/180}$; (c) ${\sigma _\theta = \pi/90}$; (d) ${\sigma _\theta = \pi/45}$.}
		\label{fig:3-2}
		\vspace{-1.5em}
	\end{figure}
	
	Fig. \ref{fig:3-2} plots the MSEs of DoA estimation versus the SNR.
	We can see that the MSEs obtained by all the methods are close to the PCRB bound when $\sigma _\theta = \pi/360$.
	When $\sigma _\theta$ increases, the curves of MSEs obtained by the PCRB oriented and PSBP-2 oriented methods get far away from the PCRB, while the curves of MSEs obtained by the PSBP-1 oriented method are still close to the PCRB.
	This indicates that the designed transmit beampattern by the proposed PSBP-1 oriented method has better DoA estimation performance than the other two methods.
	
	\vspace{-1em}
	\subsection{Discussion and Analysis}
	
	In light of the simulation results above, we further discuss the reasons behind the phenomenon that minimizing PCRB is not always effective, and compare the advantages and disadvantages of PCRB and PSBP-oriented transmit beampattern design.
	
	First, we conclude the reasons why minimizing PCRB is not always effective as follows.
	\begin{enumerate}
		\item \textbf{Theoretical Limitation:} PCRB characherizes only a theoretical lower bound. Practical estimator may not achieve this bound.
		\item \textbf{transmit beampattern Limitation:} The optimization objective may not be consistent with the actual requirements. PCRB minimization aims at mathematical optimality, but it may not guarantee sufficient energy allocation in high-probability regions, which is often critical in practice.
	\end{enumerate}
	
	Second, we compare theses transmit beampattern design approaches in terms of beampattern control flexibility, interpretability and complexity in Table \ref{tab:2}.
	
	\begin{table}[!h] 
		\vspace{-0.5em}
		\centering
		\scriptsize
		\caption{Comparison among PCRB-, PSBP-1-, and PSBP-2-Oriented Methods (The number of $\checkmark$ represents the degree.)}
		\begin{tabular}{|c|c|c|c|}
			\hline
			Aspect & PCRB Oriented & PSBP-1 Oriented & PSBP-2 Oriented \\ \hline
			\makecell{Beampattern\\ Control\\Flexibility} 
			& $\checkmark$              
			& $\checkmark \checkmark \checkmark$             
			& $\checkmark \checkmark$              
			\\ \hline
			Interpretability                
			& $\checkmark \checkmark \checkmark$           
			& $\checkmark$               
			& $\checkmark$               \\ \hline
			Complexity                      
			& $\checkmark \checkmark$            
			& $\checkmark \checkmark \checkmark$             
			& $\checkmark$               \\ \hline
		\end{tabular}\label{tab:2}
		\vspace{-1.5em}
	\end{table}

	\vspace{-1em}
	\section{Conclusion}
	
	In this paper, we studied and discussed how to exploit target location distribution for MIMO radar transmit beampattern design. 
	Specifically, we started by modeling the colocated radar system and the considered target location distribution.
	With these models at hand, the first category of target location distribution exploitation method, PCRB, was derived from the DoA estimation perspective. 
	With this as a reference, we formulated a PCRB oriented transmit beampattern design problem.
	Then, a novel PSBP based target location distribution exploitation method was proposed, inspired by the beampattern design perspective. 
	Guided by this, we formulated two PSBP oriented transmit beampattern design problems, fair PSBP design, and integrated PSBP design, for different design requirements of desirable radar beampattern.
	To solve the formulated PCRB and PSBP oriented problems, we devised the corresponding low-complexity and convergence-guaranteed algorithms.
	Finally, we provided numerical simulations in different scenarios to comprehensively evaluate the radar performance, giving the following insights:
	\begin{itemize}
		\item PCRB serves as the theoretical lower bound of DoA estimation performance; therefore, minimizing PCRB does not always result in satisfactory practical DoA estimation results. Nevertheless, PCRB has a clear physical meaning for measuring radar estimation performance.
		\item The PSBP based target location distribution exploitation can flexibly control the radar beampattern, yielding good DoA estimation performance, especially with PSBP-1. However, the PSBP-1 design is a fairness design problem with the highest computational time complexity.
		\item Both PCRB and PSBP-based target location distribution exploitation methods are important approaches. In practical radar systems, the choice between PCRB and PSBP should balance computational complexity, optimization time, and DoA estimation performance.
	\end{itemize}
	In future work, we can extend the current research to explore the application of distributed radar systems and multi-target sensing.
	
	\appendices
	
	\section{Proof of Proposition \ref{prop:1}}\label{app:A}
	\setcounter{equation}{0}
	\renewcommand*{\theequation}{A-\arabic{equation}}
	
	The FIM extracted from the received signal \eqref{eq:7a} for estimating $\bm \omega$, is given by \cite{bekkerman2006target}
	\begin{equation}\label{eq:A1}
	{{\bf{F}}_{\rm{S}}}\left( {i,j} \right) =  - \mathbb{E}\left\{ {\frac{{{\partial ^2}l\left( {{\bf{Y}}|{\bm{\omega }}} \right)}}{{\partial {\bm{\omega }}\left( i \right)\partial {\bm{\omega }}\left( j \right)}}} \right\},\;i = 1,2,3,j = 1,2,3
	\end{equation}
	
	We first calculate the partial derivations as
	\begin{align}
	&\frac{{{\partial ^2}l\left( {{\bf{Y}}|{\bf{\omega }}} \right)}}{{\partial {\theta ^2}}} =  - \frac{{2{{\left| \varsigma  \right|}^2}{\rm{Tr}}\{ {{{\bf{X}}^H}{{{\bf{\dot A}}}^H}\left( \theta  \right){\bf{\dot A}}\left( \theta  \right){\bf{X}}}\}}}{{\sigma _{\rm{r}}^2}}\label{eq:A2}\\
	&\frac{{{\partial ^2}l\left( {{\bf{Y}}|{\bf{\omega }}} \right)}}{{\partial \theta \partial {\varsigma _{\rm{R}}}}} =  - \frac{2}{{\sigma _{\rm{r}}^2}}{\rm{Tr}}\{ {{{\bf{X}}^H}{{{\bf{\dot A}}}^H}\left( \theta  \right){\bf{A}}\left( \theta  \right){\bf{X}}} \}{\varsigma _{\rm{R}}}\label{eq:A3}\\
	&\frac{{{\partial ^2}l\left( {{\bf{Y}}|{\bf{\omega }}} \right)}}{{\partial \theta \partial {\varsigma _{\rm{I}}}}} =  - \frac{2}{{\sigma _{\rm{r}}^2}}{\rm{Tr}}\{ {{{\bf{X}}^H}{{{\bf{\dot A}}}^H}\left( \theta  \right){\bf{A}}\left( \theta  \right){\bf{X}}} \}{\varsigma _{\rm{I}}}\label{eq:A4}\\
	&\frac{{{\partial ^2}l\left( {{\bf{Y}}|{\bf{\omega }}} \right)}}{{\partial \left[ {{\varsigma _{\rm{R}}},{\varsigma _{\rm{I}}}} \right]\partial {{\left[ {{\varsigma _{\rm{R}}},{\varsigma _{\rm{I}}}} \right]}^T}}} = -\frac{2}{{\sigma _{\rm{r}}^2}}{\rm{Tr}}\!\left\{ {{{\bf{X}\!}^H}\!{{\bf{A}\!}^H}\!\left( \theta  \right)\!{\bf{A}}\!\left( \theta  \right){\bf{X}}} \right\}\!{{\bf{I}}_2}\label{eq:A5}
	\end{align}
	where  ${{{\bf{\dot A}}}^H}\left( \theta  \right){\bf{\dot A}}\left( \theta  \right)    =  \left\| {{{{\bf{\dot a}}}_r}\left( \theta  \right)} \right\|_F^2{{\bf{a}}_t}\left( \theta  \right){\bf{a}}_t^H\left( \theta  \right) + {M_r}{{{\bf{\dot a}}}_t}\left( \theta  \right){\bf{\dot a}}_t^H\left( \theta  \right)$, ${{{\bf{\dot A}}}^H}\left( \theta  \right){\bf{A}}\left( \theta  \right)    =    {M_r}{{{\bf{\dot a}}}_t}\left( \theta  \right){\bf{a}}_t^H\left( \theta  \right)$, ${{\bf{A}}^H}\left( \theta  \right){\bf{A}}\left( \theta  \right) = {M_r}{{\bf{a}}_t}\left( \theta  \right){\bf{a}}_t^H\left( \theta  \right)$\footnote{Note that the center of ULA is the phase reference, leading to ${{{\bf{\dot a}}}^H}\left( \theta  \right){\bf{a}}\left( \theta  \right) = {{\bf{a}}^H}\left( \theta  \right){\bf{\dot a}}\left( \theta  \right) = 0$.}.
	Then, the elements of the FIM ${\bf F}_{\rm S}$ can be expressed as
	\begin{align}
	&{F_{\theta \theta }} = \textstyle\int_{ - \pi /2}^{\pi /2} {f\left( \theta  \right)\frac{{2{{\left| \varsigma  \right|}^2}{\rm{Tr}}\{ {{{\bf{X}}^H}{{{\bf{\dot A}}}^H}\left( \theta  \right){\bf{\dot A}}\left( \theta  \right){\bf{X}}} \}}}{{\sigma _{\rm{r}}^2}}d\theta } \nonumber\\
	&\quad\;\;=\frac{{2{{\left| \varsigma  \right|}^2}}}{{\sigma _{\rm{r}}^2}}{\rm{Tr}}\left\{ {{{\bf{X}}^H}{{\bf{\Xi }}_1}{\bf{X}}} \right\}\label{eq:A6}\\
	&{{\bf{F}}_{\theta \varsigma }}\left( {1,1} \right) = \textstyle\int_{ - \pi /2}^{\pi /2} {f\left( \theta  \right)\frac{2}{{\sigma _{\rm{r}}^2}}{\rm{Tr}}\{ {{{\bf{X}}^H}{{{\bf{\dot A}}}^H}\left( \theta  \right){\bf{A}}\left( \theta  \right){\bf{X}}} \}{\varsigma _{\rm{R}}}d\theta } \nonumber\\
	&\qquad\qquad= \frac{{2{\varsigma _{\rm{R}}}}}{{\sigma _{\rm{r}}^2}}{\rm{Tr}}\{ {{{\bf{X}}^H}{{\bf{\Xi }}_2}{\bf{X}}} \}\label{eq:A7}\\
	&{{\bf{F}}_{\theta \varsigma }}\left( {1,2} \right) = \textstyle\int_{ - \pi /2}^{\pi /2} {f\left( \theta  \right)\frac{2}{{\sigma _{\rm{r}}^2}}{\rm{Tr}}\{ {{{\bf{X}}^H}{{{\bf{\dot A}}}^H}\left( \theta  \right){\bf{A}}\left( \theta  \right){\bf{X}}} \}{\varsigma _{\rm{I}}}d\theta } \nonumber\\
	&\qquad\qquad= \frac{{2{\varsigma _{\rm{I}}}}}{{\sigma _{\rm{r}}^2}}{\rm{Tr}}\left\{ {{{\bf{X}}^H}{{\bf{\Xi }}_2}{\bf{X}}} \right\}\label{eq:A8}\\
	&{{\bf{F}}_{\varsigma \varsigma }} =\textstyle\int_{ - \pi /2}^{\pi /2}{f\left( \theta  \right)\frac{2}{{\sigma _{\rm{r}}^2}}{\rm{Tr}}\left\{ {{{\bf{X}}^H}{{\bf{A}}^H}\left( \theta  \right){\bf{A}}\left( \theta  \right){\bf{X}}} \right\}{{\bf{I}}_2}d\theta } \nonumber\\
	&\quad\;\;=\frac{2}{{\sigma _{\rm{r}}^2}}{\rm{Tr}}\{ {{{\bf{X}}^H}{{\bf{\Xi }}_3}{\bf{X}}} \}{{\bf{I}}_2}\label{eq:A9}
	\end{align}
	where ${{\bf{\Xi }}_1} = \int_{ - \pi /2}^{\pi /2} {f\left( \theta  \right)} (\left\| {{{{\bf{\dot a}}}_r}\left( \theta  \right)} \right\|_F^2{{\bf{a}}_t}\left( \theta  \right){\bf{a}}_t^H\left( \theta  \right) + {M_r}{{{\bf{\dot a}}}_t}\left( \theta  \right){\bf{\dot a}}_t^H\left( \theta  \right))d\theta $, ${{\bf{\Xi }}_2} = {M_r}\int_{ - \pi /2}^{\pi /2} {f\left( \theta  \right){{{\bf{\dot a}}}_t}\left( \theta  \right){\bf{a}}_t^H\left( \theta  \right)d\theta } $, ${{\bf{\Xi }}_3} = {M_r}\int_{ - \pi /2}^{\pi /2} {f\left( \theta  \right){{\bf{a}}_t}\left( \theta  \right){\bf{a}}_t^H\left( \theta  \right)d\theta } $.
	
	Based on \eqref{eq:A7} and \eqref{eq:A8}, we can easily obtain
	\begin{equation}\label{eq:A10}
	{{\bf{F}}_{\theta \varsigma }} = \frac{2}{{\sigma _{\rm{r}}^2}}{\rm{Tr}}\left\{ {{{\bf{X}}^H}{{\bf{\Xi }}_2}{\bf{X}}} \right\}\left[ {{\varsigma _{\rm{R}}},{\varsigma _{\rm{I}}}} \right]
	\end{equation}
	Due to the symmetry of ${\bf F}_{\rm S}$, the equality relationship ${{\bf{F}}_{\varsigma \theta }} = {\bf{F}}_{\theta \varsigma }^H$ is held.
	
	The proof is completed.
	
	\section{Proof of Proposition \ref{prop:2}}\label{app:B}
	\setcounter{equation}{0}
	\renewcommand*{\theequation}{B-\arabic{equation}}
	
	We can expand the equation \eqref{eq:15} by substituting the original expression of ${{\bf{\Xi }}_1}$, ${{\bf{\Xi }}_2}$ and ${{\bf{\Xi }}_3}$ in \eqref{eq:B1}.
	\begin{figure*}[!ht]
		\begin{equation}\label{eq:B1}
		\begin{aligned}
		&{\rm{PCR}}{{\rm{B}}_\theta }\left( {\bf{X}} \right)=\left[ {\Lambda  + \frac{{2{{\left| \varsigma  \right|}^2}}}{{\sigma _{\rm{r}}^2}}{\rm{Tr}}\left\{ {{{\bf{X}}^H}\left( {\textstyle\int_{ - \pi /2}^{\pi /2} {f\left( \theta  \right)\left\| {{{{\bf{\dot a}}}_r}\left( \theta  \right)} \right\|_F^2{{\bf{a}}_t}\left( \theta  \right){\bf{a}}_t^H\left( \theta  \right)d\theta } } \right){\bf{X}}} \right\}} \right.\\
		&\quad{\left. { + \frac{{2{{\left| \varsigma  \right|}^2}}}{{\sigma _{\rm{r}}^2}}\underbrace {\left( {{\rm{Tr}}\left\{ {{{\bf{X}}^H}( {{M_r}\textstyle\int_{ - \pi /2}^{\pi /2} {f\left( \theta  \right){{{\bf{\dot a}}}_t}\left( \theta  \right){\bf{\dot a}}_t^H\left( \theta  \right)d\theta } } ){\bf{X}}} \right\} - \frac{{{| {{\rm{Tr}}\{ {{{\bf{X}}^H}( {{M_r}\int_{ - \pi /2}^{\pi /2} {f\left( \theta  \right){{{\bf{\dot a}}}_t}\left( \theta  \right){\bf{a}}_t^H\left( \theta  \right)d\theta } } ){\bf{X}}} \}} |^2}}}{{{\rm{Tr}}\{ {{{\bf{X}}^H}( {{M_r}\int_{ - \pi /2}^{\pi /2} {f\left( \theta  \right){{\bf{a}}_t}\left( \theta  \right){\bf{a}}_t^H\left( \theta  \right)d\theta } } ){\bf{X}}} \}}}} \right)}_{q\left( {\bf{X}} \right)}} \right]^{ - 1}}
		\end{aligned}
		\end{equation}
		\begin{equation}\label{eq:B2}
		\begin{aligned}
		{\rm Nu} = \!\!\int_{ - \pi /2}^{\pi /2}\!\!\!\!\!\!\!\!{f\left( \theta  \right){\rm{Tr}}\!\left\{ {{{\bf{X}}^H}\!\!\left( {{{{\bf{\dot a}}}_t}\left( \theta  \right){\bf{\dot a}}_t^H\!\left( \theta  \right)} \right)\!\!{\bf{X}}} \right\}\!\!d\theta }\!\! \int_{ - \pi /2}^{\pi /2}\!\!\!\!\!\!\!\! {f\left( \theta  \right){\rm{Tr}}\!\left\{ {{{\bf{X}}^H}\!\!\left( {{{\bf{a}}_t}\left( \theta  \right){\bf{a}}_t^H\!\left( \theta  \right)} \right)\!\!{\bf{X}}} \right\}\!\!d\theta }  \!-\! {\left| {\int_{ - \pi /2}^{\pi /2} \!\!\!\!\!\!\!\!{f\left( \theta  \right){\rm{Tr}}\!\left\{ {{{\bf{X}}^H}\!\!\left( {{{{\bf{\dot a}}}_t}\left( \theta  \right){\bf{a}}_t^H\left( \theta  \right)} \right)\!\!{\bf{X}}} \right\}\!\!d\theta } } \right|^2}
		\end{aligned}
		\end{equation}
		\hrule
		\vspace{-1em}
	\end{figure*}
	Obviously, we can tackle ${q\left( {\bf{X}} \right)}$ in \eqref{eq:B1} by reducing fractions to a common denominator, where the common denominator is ${\rm De}={{\rm{Tr}}\{ {{{\bf{X}}^H}( {{M_r}\int_{ - \pi /2}^{\pi /2} {f( \theta ){{\bf{a}}_t}( \theta  ){\bf{a}}_t^H( \theta )d\theta } }){\bf{X}}} \}}>0$ and the numerator is given in \eqref{eq:B2}.
	Applying Cauchy-Schwarz inequality, we can obtain ${\rm Nu} \ge 0$, and then ${q\left( {\bf{X}} \right)} \ge 0$.
	Accordingly, the upper bound of the PCRB, ${\overline {{\rm{PCRB}}} _\theta }\left( {\bf{X}} \right)$, can be expressed as
	\begin{equation}
	{\rm{PCR}}{{\rm{B}}_\theta }\left( {\bf{X}} \right) \!<\! {\left[ {\Lambda  \!+ \!\frac{{2{{\left| \varsigma  \right|}^2}}}{{\sigma _{\rm{r}}^2}}{\rm{Tr}}\left\{ {{{\bf{X}}^H}{{\bf{\Xi }}_0}{\bf{X}}} \right\}} \right]^{ - 1}} \!\!\!\!\!\!\buildrel \Delta \over = {\overline {{\rm{PCRB}}} _\theta }\left( {\bf{X}} \right)
	\end{equation}
	where ${{\bf{\Xi }}_0} = {\int_{ - \pi /2}^{\pi /2} {f\left( \theta  \right)\left\| {{{{\bf{\dot a}}}_r}\left( \theta  \right)} \right\|_F^2{{\bf{a}}_t}\left( \theta  \right){\bf{a}}_t^H\left( \theta  \right)d\theta } }$.
	
	The proof is completed.
	
	\section{Proof of Proposition \ref{prop:3}}\label{app:C}
	\setcounter{equation}{0}
	\renewcommand*{\theequation}{C-\arabic{equation}}
	
	Before proving Proposition \ref{prop:3} from iterative steps \eqref{eq:26-a}-\eqref{eq:26-c}, the following optimality condition holds.
	\begin{equation}
		\begin{aligned}
		{\bf{0}}& = {\nabla _{\bf{X}}}{\cal L}{_1}({{\bf{X}}^{t + 1}},{{\bf{U}}^{t + 1}},{\bf{D}}_1^t)\nonumber\\
		&= \nabla f\left( {{{\bf{X}}^{t + 1}}} \right) - {\rho _1}\left( {{{\bf{U}}^{t + 1}} - {{\bf{X}}^{t + 1}} + {\bf{D}}_1^t} \right)\nonumber\\
		&= \nabla f\left( {{{\bf{X}}^{t + 1}}} \right) - {\rho _1}{\bf{D}}_1^{t + 1}
		\end{aligned}\label{eq:C1}
	\end{equation}
	where $f\left( {\bf{X}} \right) =  - {\rm{Tr}}\left\{ {{{\bf{X}}^H}{{\bf{\Xi }}_0}{\bf{X}}} \right\}$.
	Besides, we obtain the inequality relationship
	\begin{equation}
	\begin{aligned}
	&{\| {\nabla f\left( {{{\bf{X}}^{t + 1}}} \right) - \nabla f\left( {{{\bf{X}}^t}} \right)} \|_F} \\
	&= {\| { - \left( {{{\bf{\Xi }}_0} + {\bf{\Xi }}_0^H} \right){{\bf{X}}^{t + 1}} + \left( {{{\bf{\Xi }}_0} + {\bf{\Xi }}_0^H} \right){{\bf{X}}^t}} \|_F}\\
	&\le {\| {\left( {{{\bf{\Xi }}_0} + {\bf{\Xi }}_0^H} \right)} \|_F}{\| {{{\bf{X}}^t} - {{\bf{X}}^{t + 1}}} \|_F}
	\end{aligned}\label{eq:C2}
	\end{equation}
	Based on \eqref{eq:C1}, we have
	\begin{equation}\label{eq:C3}
	{\bf{D}}_1^{t + 1} = \frac{1}{{{\rho _1}}}\nabla f\left( {{{\bf{X}}^{t + 1}}} \right)
	\end{equation}
	Substituting \eqref{eq:C3} into \eqref{eq:C2}, we can obtain
	\begin{equation}
	\begin{aligned}
	\!\!{\| {{{\bf{D}}^{t + 1}} \!-\! {{\bf{D}}^t}} \|_F} \!=&\frac{1}{{{\rho _1}}}{\| {\nabla f\left( {{{\bf{X}}^{t + 1}}} \right) - \nabla f\left( {{{\bf{X}}^t}} \right)} \|_F}\\
	\le &\frac{1}{{{\rho _1}}}{\| {\left( {{{\bf{\Xi }}_0} + {\bf{\Xi }}_0^H} \right)} \|_F}{\| {{{\bf{X}}^t} \!-\!{{\bf{X}}^{t + 1}}} \|_F}
	\end{aligned}\label{eq:C4}
	\end{equation}
	
	Proof of (a):
	The difference of the augmented Lagrangian \eqref{eq:L1}  can be rewritten as
	\begin{equation}
	\begin{aligned}
	&{{\cal L}_1}({{\bf{X}}^{t + 1}},{{\bf{U}}^{t + 1}},{\bf{D}}_1^{t + 1}) - {{\cal L}_1}({{\bf{X}}^t},{{\bf{U}}^t},{\bf{D}}_1^t)\\
	&= \underbrace {{{\cal L}_1}({{\bf{X}}^t},{{\bf{U}}^{t + 1}},{\bf{D}}_1^t) - {{\cal L}_1}({{\bf{X}}^t},{{\bf{U}}^t},{\bf{D}}_1^t)}_{(1)}\\
	&\quad+\underbrace {{{\cal L}_1}({{\bf{X}}^{t + 1}},{{\bf{U}}^{t + 1}},{\bf{D}}_1^{t + 1}) - {{\cal L}_1}({{\bf{X}}^{t + 1}},{{\bf{U}}^{t + 1}},{\bf{D}}_1^t)}_{(2)}\\
	&\quad+\underbrace {{{\cal L}_1}({{\bf{X}}^{t + 1}},{{\bf{U}}^{t + 1}},{\bf{D}}_1^t) - {{\cal L}_1}({{\bf{X}}^t},{{\bf{U}}^{t + 1}},{\bf{D}}_1^t)}_{(3)}
	\end{aligned}\label{eq:C5}
	\end{equation}
	It is obvious that the second term (1) is bounded by 
	\begin{equation}\label{eq:C6}
	{{\cal L}_1}({{\bf{X}}^t},{{\bf{U}}^{t + 1}},{\bf{D}}_1^t) - {{\cal L}_1}({{\bf{X}}^t},{{\bf{U}}^t},{\bf{D}}_1^t) \le 0
	\end{equation}
	due to the updating of ${\bf U}^{t+1}$ in the subproblem \eqref{eq:26-a} under ADMM framework.
	
	Specifically, the first term (2) is bounded by
	\begin{align}
	&{{\cal L}_1}({{\bf{X}}^{t + 1}},{{\bf{U}}^{t + 1}},{\bf{D}}_1^{t + 1}) - {{\cal L}_1}({{\bf{X}}^{t + 1}},{{\bf{U}}^{t + 1}},{\bf{D}}_1^t)\nonumber\\
	& = \frac{{{\rho _1}}}{2}( {2\| {{\bf{D}}_1^{t + 1} - {\bf{D}}_1^t} \|_F^2 + \| {{\bf{D}}_1^{t + 1}} \|_F^2 - \left\| {{\bf{D}}_1^t} \right\|_F^2} )\nonumber\\
	& \mathop  \le \limits^{(\rm i - a)} \frac{{{\rho _1}}}{2}( {2\left\| {{\bf{D}}_1^{t + 1} - {\bf{D}}_1^t} \right\|_F^2 + \| {{\bf{D}}_1^{t + 1} - {\bf{D}}_1^t} \|_F^2} )\nonumber \\
	&= \frac{{3{\rho _1}}}{2}\| {{\bf{D}}_1^{t + 1} - {\bf{D}}_1^t} \|_F^2\nonumber\\
	& \mathop  \le \limits^{(\rm i- b)} \frac{3}{{2{\rho _1}}}\| {( {{{\bf{\Xi }}_0} + {\bf{\Xi }}_0^H} )} \|_F^2\| {{{\bf{X}}^{t + 1}} - {{\bf{X}}^t}} \|_F^2
	\end{align}\label{eq:C7}%
	where (i-a) holds due to the triangle inequality, and (i-b) holds due to \eqref{eq:C4}.

	The term (3) is bounded by
	\begin{align}
	&{{\cal L}_1}({{\bf{X}}^{t + 1}},{{\bf{U}}^{t + 1}},{\bf{D}}_1^t) - {{\cal L}_1}({{\bf{X}}^t},{{\bf{U}}^{t + 1}},{\bf{D}}_1^t)\nonumber\\
	& = f( {{{\bf{X}}^{t + 1}}} ) - f( {{{\bf{X}}^t}} ) + \frac{{{\rho _1}}}{2}\| {{{\bf{U}}^{t + 1}} + {\bf{D}}_1^t - {{\bf{X}}^{t + 1}}} \|_F^2 \nonumber\\
	&\quad- \frac{{{\rho _1}}}{2}\| {{{\bf{U}}^{t + 1}} + {\bf{D}}_1^t - {{\bf{X}}^t}} \|_F^2\nonumber\\
	&\mathop  = \limits^{(\rm ii - a)} f( {{{\bf{X}}^{t + 1}}} ) - f( {{{\bf{X}}^t}} ) - \frac{{{\rho _1}}}{2} {\| {{{\bf{X}}^{t + 1}} - {{\bf{X}}^t}} \|_F^2} \nonumber\\
	&\qquad{ + 2\Re \{ {{\rm{Tr}}\{ {{{( {{{\bf{U}}^{t + 1}} - {{\bf{X}}^{t + 1}} + {\bf{D}}_1^t} )}^H}( {{\bf{X}}_1^t - {\bf{X}}_1^{t + 1}} )} \}} \}}\nonumber\\
	& = f( {{{\bf{X}}^{t + 1}}} ) - f( {{{\bf{X}}^t}} ) - \frac{{{\rho _1}}}{2}\left\| {{{\bf{X}}^{t + 1}} - {{\bf{X}}^t}} \right\|_F^2 \nonumber\\
	&\quad- {\rho _1}\Re \{ {{\rm{Tr}}\{ {{{( {{\bf{D}}_1^{t + 1}} )}^H}( {{\bf{X}}_1^{t + 1} - {\bf{X}}_1^t} )} \}} \}\nonumber\\
	&\mathop  = \limits^{(\rm ii - b)} - \frac{{{\rho _1}}}{2}\| {{{\bf{X}}^{t + 1}} - {{\bf{X}}^t}} \|_F^2 - ( {f( {{{\bf{X}}^t}} ) - f( {{{\bf{X}}^{t + 1}}} )} )\nonumber\\
	&\qquad + \Re \{ {{\rm{Tr}}\{ {{{( {\nabla f\left( {{{\bf{X}}^{t + 1}}} \right)} )}^H}( {{\bf{X}}_1^t - {\bf{X}}_1^{t + 1}} )} \}} \} \nonumber\\
	&\mathop  \le \limits^{(\rm ii - c)}  - \frac{{{\rho _1}}}{2}\| {{{\bf{X}}^{t + 1}} - {{\bf{X}}^t}} \|_F^2	    
	\end{align}\label{eq:C8}
	where (ii-a) holds due to the cosine equality: ${\left\| {{\bf b} + {\bf c}} \right\|^2} - {\left\| {{\bf a} + {\bf c}} \right\|^2} = {\left\| {{\bf b} - {\bf a}} \right\|^2} + 2\left\langle {{\bf a} + {\bf c},{\bf b} - {\bf a}} \right\rangle$ by substituting ${\bf a} = {{\rm vec}\left({\bf{X}}^{t + 1}\right)},{\bf b} = {{\rm vec}\left({\bf{X}}^t\right)},{\bf c} =  - {{\rm vec}\left({\bf{U}}^{t + 1}\right)} - {\rm vec}\left({\bf{D}}_1^t\right)$.
	(ii-b) holds due to \eqref{eq:C3}, and (ii-c) holds due to the property of concave function $f\left( {\bf{X}} \right)$.
	
	Applying \eqref{eq:C6}-(\hyperref[eq:C8]{C-7}) into \eqref{eq:C5}, we can obtain
	\begin{equation}
	\begin{aligned}
	&{{\cal L}_1}({{\bf{X}}^{t + 1}},{{\bf{U}}^{t + 1}},{\bf{D}}_1^{t + 1}) - {{\cal L}_1}({{\bf{X}}^t},{{\bf{U}}^t},{\bf{D}}_1^t)\\
	&\le ( {\frac{3}{{2{\rho _1}}}\| {( {{{\bf{\Xi }}_0} + {\bf{\Xi }}_0^H} )} \|_F^2 - \frac{{{\rho _1}}}{2}} )\| {{{\bf{X}}^{t + 1}} - {{\bf{X}}^t}} \|_F^2
	\end{aligned}\label{eq:C9}
	\end{equation}
	which means that the value of the augmented Lagrangian \eqref{eq:L1} is always decreasing during the iterative process when the penalty parameter ${\rho _1} > \sqrt 3 {\| { {{{\bf{\Xi }}_0} + {\bf{\Xi }}_0^H} }\|_F}$.
	
	Proof of (b):
	Based on \eqref{eq:P1-3}, ${\bf X}^{t+1}$ is bounded by the constraint $\| {\bf{X}}^{t+1} \|_F^2 = P$, and $f( {\bf{X}}^{t+1} )$ is continuous, so $f( {\bf{X}}^{t+1} )$ is bounded on $[f_{\rm min},f_{\rm max}]$, where $f_{\rm min}$ is the lower bound of $f( {\bf{X}}^{t+1} )$ and $f_{\rm max}$ is the upper bound of $f( {\bf{X}}^{t+1} )$.
	Since $\| {\bf{U}}^{t+1} - {\bf{X}}^{t+1} + {{\bf{D}}_1}^{t+1} \|_F^2 \ge 0$, we have $ {{\cal L}_1}({\bf{X}}^{t+1},{\bf{U}}^{t+1},{\bf{D}}^{t+1}_1) \ge f_{\rm min}$.
	Therefore, ${\cal L}{_1}({{\bf{X}}^{t+1},{\bf{U}}^{t+1},{\bf{D}}^{t+1}_1})$ is lower bounded for all $t$ and converges as $t \to  + \infty$ based on (a).
	
	Proof of (c): 
	Based on (a) and (b), setting ${\rho _1} > \sqrt 3 {\| { {{{\bf{\Xi }}_0} + {\bf{\Xi }}_0^H} }\|_F}$, we have
	\begin{equation}\label{eq:C10}
	\mathop {\lim }\limits_{t \to \infty } \| {{{\bf{X}}^{t + 1}} - {{\bf{X}}^t}} \|_F^2 = 0
	\end{equation}
	Applying \eqref{eq:C4}, we have
	\begin{equation}\label{eq:C11}
	\mathop {\lim }\limits_{t \to \infty } \| {{{\bf{D}}^{t + 1}} - {{\bf{D}}^t}} \|_F^2 = 0
	\end{equation}
	Substituting \eqref{eq:26-c} into \eqref{eq:C11}, we can obtain
	\begin{equation}\label{eq:C12}
	\mathop {\lim }\limits_{t \to \infty } \| {{{\bf{U}}^{t + 1}} - {{\bf{X}}^{t + 1}}} \|_F^2= 0
	\end{equation}
	which proves (c).
	
	The proof is completed.
	
	\balance
	\bibliographystyle{IEEEtran}
	\bibliography{IEEEabrv,ref_abrv.bib}

 \end{document}